\pdfoutput=1
\documentclass{article}
\usepackage{arxiv}
\usepackage[utf8]{inputenc} 
\usepackage[T1]{fontenc}    
\usepackage{hyperref}       
\usepackage{url}            
\usepackage{booktabs}       
\usepackage{amsfonts}       
\usepackage{nicefrac}       
\usepackage{microtype}      
\usepackage{lipsum}
\usepackage{amsmath,amssymb,bbold}
\usepackage{mathrsfs}
\usepackage{cite}
\usepackage{nameref,hyperref}
\usepackage{microtype}
\DisableLigatures[f]{encoding = *, family = * }
\usepackage[table]{xcolor}
\usepackage{array}
\newcolumntype{+}{!{\vrule width 2pt}}
\newlength\savedwidth

\usepackage[aboveskip=1pt,labelfont=bf,labelsep=period,justification=raggedright,singlelinecheck=off]{caption}

\usepackage{subfig}
\usepackage{tikz}
\usepackage{empheq}
\usepackage{pgfplots}
\usepackage{pgfplotstable}
\usepackage{overpic}
\usepackage{url}
\usepackage{psfrag}

\def\Bchi{\mbox{\boldmath$\chi$}}

\def\Bvarphi{\mbox{\boldmath$\varphi$}}
\def\bone{\mbox{\boldmath$1$}}
\def\bzero{\mbox{\boldmath$0$}}

\def\bF{\mbox{\boldmath$ F$}}

\def\bN{\mbox{\boldmath$ N$}}

\def\bP{\mbox{\boldmath$ P$}}

\def\bX{\mbox{\boldmath$ X$}}

\def\bs{\mbox{\boldmath$ s$}}

\def\bu{\mbox{\boldmath$ u$}}
\def\bv{\mbox{\boldmath$ v$}}

\def\bx{\mbox{\boldmath$ x$}}

\newcommand{\red}[1]{\textcolor{black}{#1}}
\newcommand{\strike}[1]{}

\pgfplotsset{compat=1.13}
\begin{document}
\vspace*{0.2in}

\begin{flushleft}
{\Large
\textbf\newline{A computational framework for the morpho-elastic development of molluskan shells by surface and volume growth} 
}
\newline
\\
Shiva Rudraraju \textsuperscript{1},
Derek E. Moulton \textsuperscript{2},
R\'egis Chirat \textsuperscript{3},
Alain Goriely\textsuperscript{2},
Krishna Garikipati \textsuperscript{4*}
\\
\bigskip 
\textbf{1} Department of Mechanical Engineering, University of Wisconsin-Madison, Wisconsin, USA.
\\
\textbf{2} Mathematical Institute, University of Oxford, Oxford, UK.
\\
\textbf{3}  Université Lyon1, CNRS UMR 5276 LGL-TPE, France.
\\
\textbf{4}  Departments of Mechanical Engineering and Mathematics, University of Michigan, Ann Arbor, Michigan, USA. *Corresponding author. Email: krishna@umich.edu.
\\

\end{flushleft}
\section*{Abstract}
Mollusk shells are an ideal model system for understanding the morpho-elastic basis of morphological evolution of invertebrates' exoskeletons. During the formation of the shell, the mantle tissue secretes proteins and minerals that calcify to form a new incremental layer of the exoskeleton. Most of the existing literature on the morphology of mollusks is descriptive. The mathematical understanding of the underlying coupling between pre-existing shell morphology, \emph{de novo} surface deposition and morpho-elastic volume growth is at a nascent stage, primarily limited to reduced geometric representations. Here, we propose a general, three-dimensional computational framework coupling pre-existing morphology, incremental surface growth by accretion, and morpho-elastic volume growth. We exercise this framework by applying it to explain the stepwise morphogenesis of seashells during growth: new material surfaces are laid down by accretive growth on the mantle whose form is determined by its morpho-elastic growth. Calcification of the newest surfaces extends the shell as well as creates a new scaffold that constrains the next growth step.  We study the effects of surface and volumetric growth rates, and of previously deposited shell geometries on the resulting modes of mantle deformation, and therefore of the developing shell's morphology. Connections are made to a range of complex shells ornamentations.

\section*{Introduction}
With around 200,000 living species, molluska are the second most diversified phylum of the animal kingdom, including gastropods (snails, slugs), bivalves (mussels, oysters,...), cephalopods (squids, \textit{Nautilus},...) and five other classes \cite{PonderLinderg2008} occupying a wide range of marine, freshwater, and terrestrial habitats. The huge morphological diversity among classes makes mollusks particularly interesting from an evolutionary perspective, notably with regard to questions related to the origin, evolution, and disparity of their body plan and their shell \cite{wanninger2015evolutionary,sigwart2017zoology}. The evolutionary success of mollusks, spanning over 540 million years, can be at least partly attributed to the shell that provides both protection and support to the soft body \cite{runnegar1996early}. Beyond their obvious aesthetic appeal, molluskan shells are an important research area in different fields. They have become exemplar model systems for studying the processes of biomineralization, a topic attracting a great deal of interest: from materials science to biomedical applications \cite{berland2005nacre,espinosa2011tablet}. Recent studies have begun to identify genes involved in these complex processes and to analyse how they are developmentally regulated \cite{jackson2007dynamic}, although the physical mechanisms underlying the morphogenesis of the shell ultrastructures remain poorly understood \cite{zlotnikov2017thermodynamic}. Recent attention has also been given to the formation and differentiation of the shell-secreting mantle margin during development \cite{audino2015mantle} and its morphological variations among classes \cite{mcdougall2011ultrastructure,westermann2005functional}. Detailed microscopic studies continue to provide important details about the structure and mutual relationships between the mantle, periostracum, and shell \cite{checa2014early}. However, despite their importance to many fields, the morphogenetic processes underlying the diversity of shapes remain elusive. This poor state of knowledge may lead to an incomplete, if not a distorted, view of the mechanisms underlying their morphological evolution. 

Several interesting theories have addressed the formation of pigmentation patterns. However, these theoretical models invoking either reaction-diffusion chemical systems \cite{meme82} or nervous activity in the mantle epithelial cells \cite{boettiger2009neural} cannot, by themselves, explain the emergence of three-dimensional forms that are subject to forces during the organism's development and life span. Indeed, while colour patterns on surfaces are primarily of biochemical origin, the formation of three-dimensional ornamentations such as ribs, tubercles, and spines is mostly a mechanical problem resulting from force generation on the mantle during growth, and its distortion in response to the force. Early studies have also considered the role of mechanics in the development of molluskan shells \cite{checa1994model,checa2002mechanics,checa2003rib,hammer2000theory,morita1991finite}. More recently, some of the authors have developed a general framework of mollusk shell morphogenesis based on continuum theories of growth and mechanics~\cite{amatar10,goriely2017mathematics}. These models have been used to study the development and evolution of shell shape from a biophysical perspective \cite{chirat2013mechanical,erlich2016morphomechanics,erlich2018mechanical,moulton2012mechanical,moulton2015morpho}. In particular, these morpho-mechanical models suggest that three-dimensional ornamentations, either parallel (i.e. commarginal ribs) or orthogonal (i.e. antimarginal ribs) to the growth lines do not require prefiguring patterns at the molecular level but may emerge \emph{de novo} from the balance of mechanical stresses intrinsic to the secreting system constrained in its growth by the calcified shell edge to which it adheres. 

Following these simplified models, we present a fully three dimensional numerical framework to study the accretive growth and nonlinear elastic deformations of the secreting mantle. As a first application, we study the effect of the geometry of the calcified shell edge, surface growth rate and morpho-elastic growth rate of the mantle on the resulting elastic deformation modes. We next study how these parameters may interact during shell development to generate diverse forms. Our main motivation for focusing on generic physical processes involved in development is that they may shape living beings in a predictive way and partly determine the spectrum of forms that have been and could have been generated during evolution. This outlook can be traced back 100 years to the pioneering work of D'Arcy Thompson, whose 1917 tome ``On Growth and Form'' \cite{thompson1942growth} continues to inspire a growing community of researchers in various fields of theoretical, evolutionary and developmental biology (e.g. \cite{saunders2017mod,goriely2017mathematics}). In this perspective, computational models of morphogenesis constitute an important tool to uncover the non-contingent rules that physical processes introduced in the development and evolution of forms.

\subsection*{Mollusk shell growth mechanics}

Molluskan shells grow via an accretive process occurring at the shell margin by an organ called the mantle, which is a thin elastic membrane lining the inner surface of the shell. Over each increment, the mantle extends slightly beyond the calcified shell edge, while adhering to the rigid shell. The mantle then secretes matrix proteins, which, through biomineralization and calcification harden into a new layer of shell. 

\begin{figure}[!ht]
\begin{center}
\includegraphics[width=0.8\textwidth]{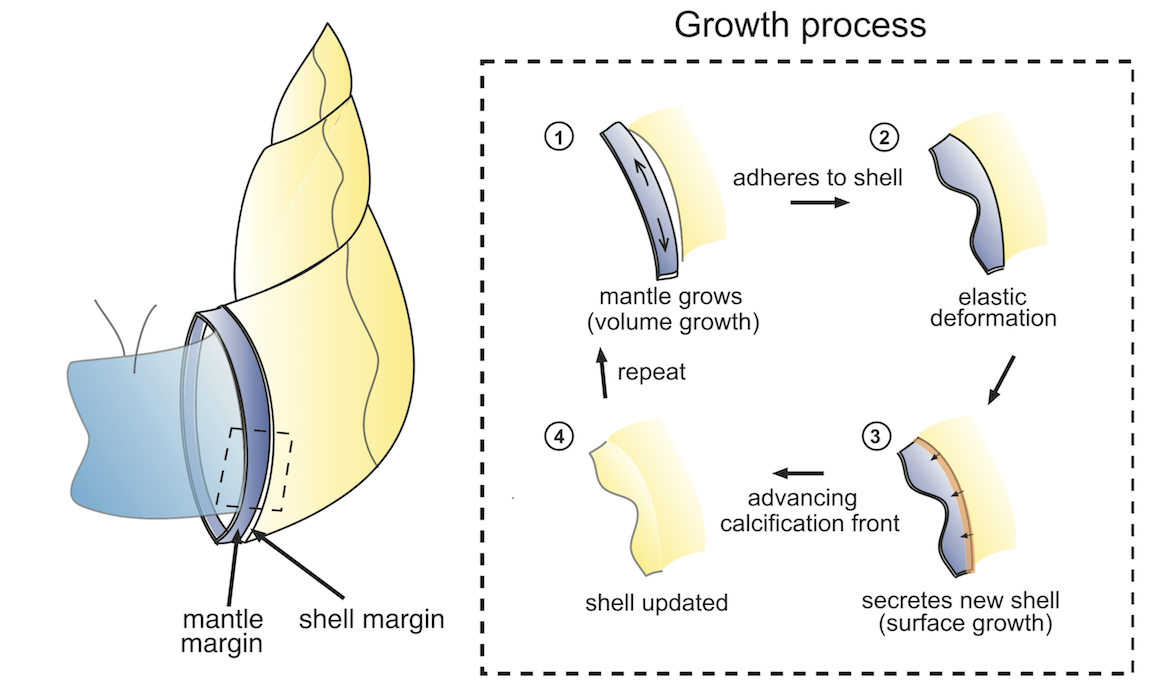}
\caption{{\bf Schematic of shell growth.} Growth process of a molluskan shell surface depicted through the steps of  volume growth of mantle tissue, morphoelastic deformation, and shell surface growth via secretion and calcification. The calcified region of the shell is indicated in yellow.}
\label{fig:ShellGrowthSchematic} 
\end{center}
\end{figure}

Within this process is an interesting mechanical interaction, due to the fact that the form taken by the mantle along the growing shell front is fixed in the calcified edge, while the form of the calcified edge partly determines the shape of the mantle at the next growth increment \cite{moulton2012mechanical}. As the mantle may have grown since the last shell secretion, its margin may be longer than the shell edge, and hence attachment to the shell may induce deformation of the mantle tissue that is then recorded and fixed in the shell shape upon secretion and calcification. We introduce our foundational notions of two distinct modes of growth. A process that creates new surface where none existed before is labelled as \emph{surface growth}. In contrast, if growth takes place by deposition where material already exists, i.e., over pre-existing volume, followed by elastic relaxation, the effect is to locally increase the material volume without adding new surface. For this reason we label it as \emph{volume growth}.
From a mechanical point of view, shell growth may thus be summarized by the steps illustrated schematically in Fig. \ref{fig:ShellGrowthSchematic}: (1) the mantle extends beyond the shell edge while also growing along the shell margin (volume growth), and (2) adhering to the rigid shell, creating an elastic deformation (morphoelasticity); in this deformed configuration, (3) new shell material is secreted (surface growth of the shell), and thus (4) a new layer of shell appears in the shape of the deformed mantle, which undergoes biomineralization and calcification, and the process repeats. 

The same basic process occurs in all shell-building mollusks, and yet produces a hugely diverse output of shell shapes and ornamentations. A general goal is to produce a mathematical and computational framework to explore this diversity: in particular how mechanical properties of the mantle, growth rates, and geometry conspire to produce the beautiful and varied outcomes observed in this phylum. However, a complete mathematical description is inherently challenging, as it links complex shell geometry (helicospiral, e.g.), elements of both surface and volume growth, nonlinear elastic deformations, and calcification. Previous work by some of the authors has approached this problem in a setting of one-dimensional elasticity, treating the interaction between the mantle margin and shell as a rod on an evolving elastic foundation. Here, our objective is to develop an algorithmic approach and computational tools to model the problem in a setting of three-dimensional nonlinear morphoelasticity. 

However, the mathematical details of such a computation are quite complex; indeed the combination of surface and volume growth is itself a significant challenge in biomechanics, with evolving reference configurations and multiple growth tensors. Here, we have the added complexity of the distinction between the growing shell and the growing mantle, as well as the additional process of shell calcification. Hence, a proper description involves the delicate treatment of surfaces evolving due to combined mechanisms of growth, mechanical forces, and a calcification front that plays the role of a moving boundary condition. To simplify the description, in this paper we formulate a mathematical description of the process that treats the mantle and shell edge as a single elastic object undergoing surface growth, volume growth, and calcification. The mathematical details are provided in the following subsection (which may be skipped by the reader whose interest lies primarily in the outcome of applying the growth models). In short, our algorithmic approach is to execute the process shown in Fig \ref{fig:ShellGrowthSchematic}. In step (1) of the figure, extension of the mantle edge beyond the shell edge is modelled as surface growth. Volume growth is assumed to occur only in the direction parallel to the shell edge, and produces an excess of length of mantle relative to the calcified shell edge. Upon attachment, the calcified shell edge acts as a boundary condition constraining one edge of the mantle margin in step (2). Elastic deformation is determined via mechanical equilibrium within the framework of finite elasticity (and computed via the finite element method). This is the morphoelastic volume growth step that causes out-of-plane deformation of the mantle. Subsequent secretion of new shell material then occurs over the extended and deformed mantle in step (3) and the calcification front advances. Configurations of both the mantle and the shell are updated in step (4), and the process is repeated. As discussed further below, there are variations possible on exactly how this process is implemented in a computational setting. Our objective in this paper is not to exhaustively explore all possibilities, but rather to demonstrate the general validity of the algorithm and examine some basic properties of the shell patterns that emerge as output.

\section*{Methods}
\subsection*{Mathematical framework} 

\begin{figure}[!ht]
\begin{center}
\psfrag{a}{$\Gamma_{\tau_1}$}
\psfrag{b}{$\Gamma_{\tau_2}$}
\psfrag{u}{\Large $\bs_2$}
\psfrag{v}{\Large $\bs_1$}
\psfrag{w}{\Large $\bs_3$}
\includegraphics[width=0.25\textwidth]{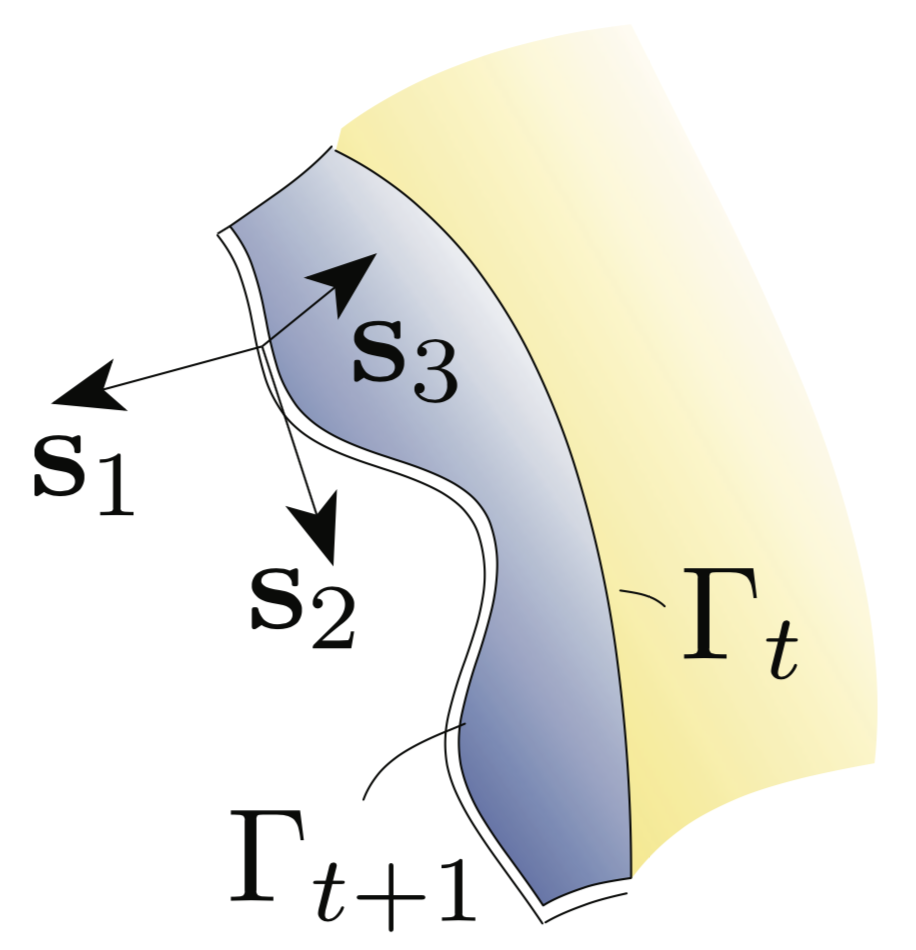}
\caption{{\bf Local coordinate system on the surface patch:}  $\bs_1$ is the direction of surface growth due to mantle extension, $\bs_2$ is the direction of volume growth along the mantle margin, and $\bs_3 = \bs_1\times \bs_2$ is the normal to the local surface patch. The calcified shell edge after time $\tau_1$ forms the generating curve $\Gamma_{\tau_1}$ for the time step, $[\tau_1,\tau_2]$; the leading edge of the grown and deformed mantle strip then forms the generating curve, $\Gamma_{\tau_2}$ for the next time step.}
\label{fig:manifoldPatch} 
\end{center}
\end{figure}

We  detail the mathematical framework that is the foundation for the eventual computational treatment of the growth processes outlined above. This requires descriptions of surface and volume growth, elasticity and calcification. 
In our model, \emph{de novo} shell material is configured by a combination of mechanisms among those introduced above: (a) surface growth (creation of new surface) by mantle extension along a unit vector, $\bs_1$, which is tangential to the shell surface and perpendicular to the mantle margin, i.e. $\bs_1$ denotes the general direction of shell growth; (b) growth in size of the mantle over its pre-existing extent (therefore ``volume'' growth) manifesting in its expansion along another unit vector $\bs_2$, which is tangential to the shell surface and in the direction of the mantle margin and shell edge \red{(since the mantle takes on the shape of the shell surface near its leading edge, the mantle margin tracks the shell edge)};  and (c) formation of crests and valleys nominally perpendicular to the undeformed shell surface, and along the unit vector $\bs_3$. We have $\bs_3 = \bs_1\times \bs_2$, and more specifically, $\bs_i\cdot\bs_j = \delta_{ij}$, where the triad $\{\bs_1,\bs_2,\bs_3\}$ changes along the curved shell surface (Fig~\ref{fig:manifoldPatch}). The third mechanism above arises from elastic bifurcations from a smooth surface, and post-bifurcation deformation driven by ``excess'' mantle  growth relative to the previous shell increment. As explained in the \red{Mollusk shell growth mechanics subsection of the Introduction}, this is the morphoelastic mechanism, which is susceptible to a continuum mechanical treatment. It is key to development of the elaborate, antimarginal decorations of molluskan shells \cite{rohomc94,amatar10,bego05,ga09,gobe05,goriely2017mathematics}. Although surface growth due to mantle extension perpendicular to the shell edge occurs only following volume growth along the shell edge, we have described the steps (a-c) in the order of surface growth, volume growth and morphoelastic deformation. This is for mathematical purposes only. In our model, these mechanisms are not separated in time. At the outset we remark that the need for precision in describing the array of configurations and mechanisms leads to complexity of notation. 

\subsubsection*{Surface growth of the mantle}

The mid-surface of the shell is represented by the surface, $\mathscr{S}_\tau \subset\mathbb{R}^3$. We regard $\mathscr{S}_\tau$ as a one-parameter family of surfaces, generated by $\tau \in [0,T]$, from a reference surface $\mathscr{S}_0 \subset\mathbb{R}^3$. The generating curve, $\Gamma_\tau \subset\partial\mathscr{S}_\tau$, is the leading edge of $\mathscr{S}_\tau$ and evolves along $\bs_1$ (see Fig~\ref{fig:manifoldPatch}). For points $\bX(\tau) \in \Gamma_t$ and $\bX(0) \in \Gamma_0$, where $\Gamma_0 \subset\partial\mathscr{S}_0$ is the initial generating curve at time $\tau = 0$, we have $\bX(\tau) = \Bchi_\tau(\bX(0))$.  Surface growth occurs by extension of the mantle along the boundary curve $\Gamma_\tau$, which advances with the velocity $\dot{\Bchi} = v_1\bs_1$. In our computational studies, we will consider spatial and temporal variations in $v_1$. \red{In principle, $v_1$ depends on space and time through quantities such as the density, stress, and chemical fields, among others, but we neglect such details in this preliminary communication.} We approximate the shell and the extended mantle as maintaining a constant thickness along $\bs_3$ throughout the growth process.

\subsubsection*{Volume growth of the mantle}
 We next consider volume growth of the mantle by expansion along $\bs_2$, which is also the tangent to $\Gamma_\tau$ \red{(both surface and volume growth of the mantle as described are a consequence of growth in size of the mollusk. It is for purposes of mathematical modelling that we have distinguished the process into surface and volume growth.)}. Due to this mechanism of growth the arc length of the fully relaxed mantle increases over time. Our treatment is focused on the kinematic manifestation of possibly inhomogeneous volume growth along $\bs_2$. We adopt the framework of finite strain elasticity, with one important variation on the traditional theme: The reference configuration of a material point is determined by its deposition time. A family of reference surfaces (2-manifolds in $\mathbb{R}^3$) is defined: $\omega_{0_\tau} = \Gamma_\tau\times (-h/2,h/2) \subset \mathbb{R}^3$, parameterized by the time of deposition, $\tau$. A material point lies on a reference surface, $\bX(\tau) \in \omega_{0_\tau}$ if it was deposited at time $\tau$. The point-to-point map of material points $\bX(\tau)$ from the reference surface $\omega_{0_\tau}$ to $\bx(t;\tau)$ on the current surface, $\omega_{t_\tau}$, is $\bx(t;\tau) = \Bvarphi(\bX(\tau),t) = \bX(\tau) + \bu(\bX(\tau),t)$, where $\bu$ is the displacement field. Note that $\omega_{t_\tau}\in \mathbb{R}^3$ also is a 2-manifold. The primary strain quantity is the deformation gradient, defined as $\bF(\bX(\tau),t)=\partial \Bvarphi(\bX(\tau),t)/\partial \bX$.  Morpho-elastic growth of the soft mantle tissue is modeled by the multiplicative decomposition of the deformation gradient $\bF(\bX(\tau),t)=\bF^\text{e}(\bX(\tau),t)\bF^\text{g}(\bX(\tau),t)$ into elastic and growth components, respectively ~\cite{gaargr04,ga09,amatar10,goriely2017mathematics}. 

\begin{figure}[!ht]
\begin{center}
\psfrag{a}{$\varphi$}
\psfrag{b}{$\bF=\bF^\text{e}\bF^\text{g}$}
\psfrag{c}{$\bF^\text{g}$}
\psfrag{d}{$\bF^\text{e}$}
\psfrag{e}{\small{Mantle with growth}}
\psfrag{f}{\small{\shortstack{Mantle with growth \\and elastic relaxation}}}
\psfrag{g}{\small{\shortstack{Mantle \\reference state}}}
\includegraphics[width=0.6\textwidth]{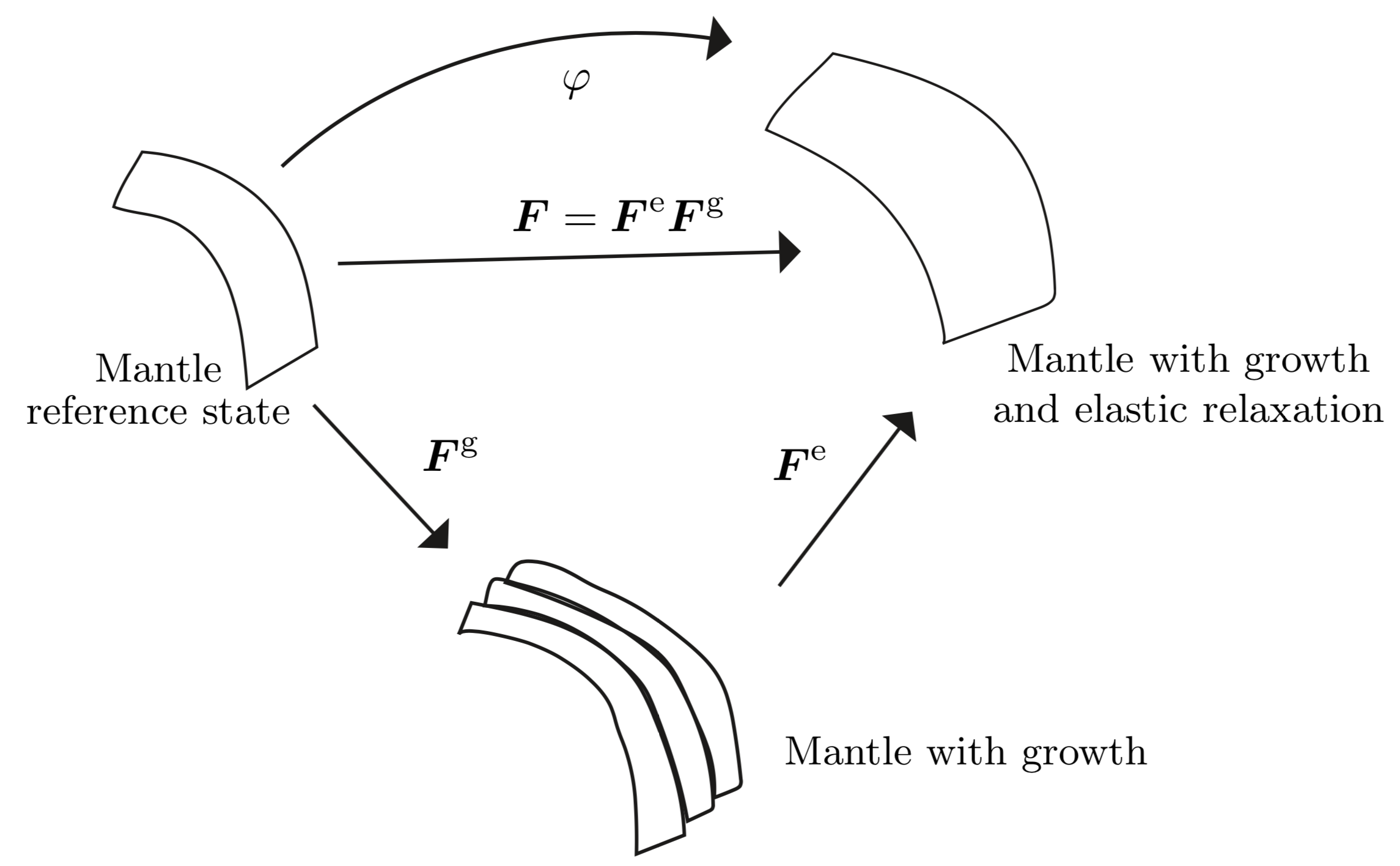}
\caption{{\bf The kinematics of growth.} The observed deformation gradient, $\bF$, is composed of an incompatible growth component, $\bF^\text{g}$, and another incompatible, but elastic component, $\bF^\text{e}$, which restores compatibility of $\bF$.}
\label{fig:kinematics} 
\end{center} 
\end{figure}

The intuitive idea is that with the mid-surface of the shell being represented by $\mathscr{S}_t$, the mantle's ``preferred'' state is given by the growth tensor $\bF^\text{g}$ relative to $\mathscr{S}_t\times (-h/2,h/2)$. Because of its attachment to the rigid shell the mantle cannot attain $\bF^\text{g}$, but only $\bF$, with $\bF^\text{e}$ being the elastic incompatibility. This multiplicative decomposition of the kinematics is the framework of morphoelasticity. It depends on the time of deposition of material points, and therefore on evolving reference configurations, and is depicted in Fig~\ref{fig:kinematics}.

In the above parametrizations $t \ge \tau$ is understood. In what follows, we will suppress functional and parameteric dependencies wherever there is no danger of confusion. 

As expressed above, our key kinematic assumption on volume growth of the mantle is that it occurs only along $\bs_2$, so that, accounting  for the appropriate tangent spaces between which $\bF^\text{g}$ is imposed,
\begin{equation}
    \dot{\bF}^\text{g} = \varepsilon_2\bF^\text{g}\bs_2 \otimes \bs_2.
    \label{Eq:rateLg}
\end{equation}

\noindent Here, $\varepsilon_2$ is the rate of the growth strain along $\bs_2$. As with the surface growth velocity, we will consider spatial and temporal variations in $\varepsilon_2$.

\subsubsection*{Secretion of shell material}
The scaffold for \emph{de novo} deposition of shell material is the mantle that has undergone an increment of surface and volume growth. New shell material is secreted on the mantle's outer surface. 

\subsubsection*{The calcification front}
While the mantle can be treated as an elastic solid, the calcified shell itself is rigid. An advancing calcification front, $\mathscr{C}_\tau \in \Bvarphi(\mathscr{S}_\tau)$, is the interface between the rigid shell and material recently secreted by the mantle. The velocity of $\mathscr{C}_\tau$ is $\bv^\text{c}$, which lies in the plane defined by $\{\bs_1,\bs_2\}$. 

\subsection*{Algorithmic formulation and implementation}

\begin{figure}[!ht]
 \begin{center}
 \psfrag{a}{$\Gamma_{t_k}$}
 \psfrag{b}{$\omega_{t_{t_k}}$}
 \psfrag{d}{$\Omega^\text{m}_{t_{t_k}}$}
 \psfrag{f}{$\Omega^\text{m}_{0_{t_{k+1}}}$}
 \psfrag{e}{$\omega_{0_{t_{k+1}}}$}
 \psfrag{i}{\small{\shortstack[l]{$\dot{\bF}^\text{g}$ imposed on $\Omega^\text{m}_{0_{t_{k+1}}}$ \\for $\Delta t$ (volume growth)}}}
 \psfrag{k}{$v_1\Delta t$}   
 \psfrag{l}{\small{\shortstack[l]{$\omega_{t_{t_k}}$ extends $v_1\Delta t\bs_1$ \\ creating $\Omega^\text{m}_{0_{t_{k+1}}}$ ({surface growth})}}}
 \psfrag{m}{$\bu=0$}
 \psfrag{g}{$\Omega^\text{m}_{t_{t_{k+1}}}$}
 \psfrag{z}{\small{\shortstack[l]{secretion and \\calcification at $t_{k+1}$}}}
 \psfrag{u}{$\bs_2$}
 \psfrag{v}{$\bs_1$}
 \psfrag{w}{$\bs_3$}
 \includegraphics[width=0.9\textwidth]{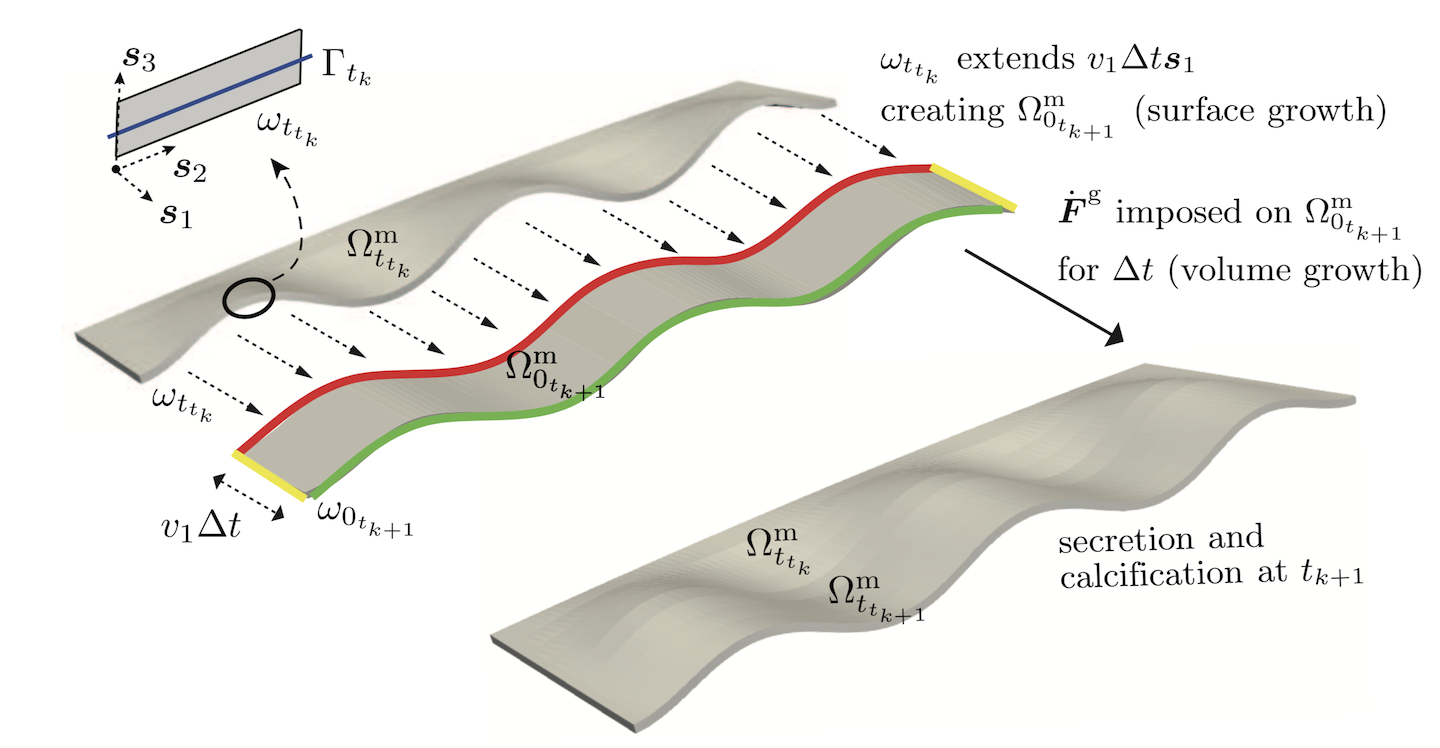}
 \caption{{\bf Mathematical model of molluskan shell growth over $(t_k,t_{k+1}]$ through a sequence of surface growth, volume growth, secretion and calcification.} Beginning with a calcified mantle current configuration $\Omega^\text{m}_{t_{t_{k}}}$, growth over $(t_k,t_{k+1}]$ is modelled via the following sequence of steps: (1) Surface growth - leading surface $\omega_{t_{t_k}}$ is displaced due to mantle extension by $ v_1\Delta t\bs_1$ to a new reference surface $\omega_{0_{t_{k+1}}}$ defining a strip of the mantle in its reference configuration $\Omega^\text{m}_{0_{t_{k+1}}}$. (2) Volume (morphoelastic) growth - $\dot{\bF}^\text{g}$ is imposed on the mantle strip $\Omega^\text{m}_{0_{t_{k+1}}}$ over $\Delta t= t_{k+1} - t_k$, driving its nonlinear deformation into the current configuration $\Omega^\text{m}_{t_{t_{k+1}}}$. (3) Shell growth occurs by secretion on the mantle strip $\Omega^\text{m}_{t_{t_{k+1}}}$, followed by (4) calcification of the mantle strip $\Omega^\text{m}_{t_{t_{k+1}}}$ at $t_{k+1}$. During the volume growth of the mantle strip from its reference configuration  $\Omega^\text{m}_{0_{t_{k+1}}}$ to its current configuration $\Omega^\text{m}_{t_{t_{k+1}}}$, boundary conditions are applied on the trailing surface (highlighted in red) and the lateral surfaces (highlighted in yellow) of the mantle strip. The front surface of the mantle strip is highlighted in green.}
 \label{fig:fronts}
 \end{center}  
\end{figure}

The first step toward an algorithmic implementation is a discretization of the continuous processes of  surface growth, morphoelastic volume growth, and evolution of the calcification front. The time interval of interest $t \in [0,T]$ is discretized by instants $t_0, t_1,\dots, t_N$, into sub-intervals $[t_0,t_1],\dots,[t_{N-1},t_N]$, where $t_0 = 0$ and $t_N = T$. For simplicity, we also consider deposition times $\tau = t_0,t_1,\dots$. In a time step $\Delta t = t_{k+1} - t_k$, the leading surface of material secreted by the mantle, $\omega_{t_t}$,  advances by $v_1\Delta t\bs_1$. Fig~\ref{fig:fronts} depicts the relevant geometry (generating curve, mantle front surface in reference and deformed configurations parameterized by deposition time) and the growth processes driving the mantle's shape by surface growth and morphoelastic volume growth. Secretion of shell material and calcification are implied, but not shown.

The preceding mathematical model is continuous in time. It describes the biological processes in the sequence of (1) the mantle's surface growth (extension), (2) morphoelastic volume growth, (3) shell growth by secretion on the mantle's current configuration, followed by (4) calcification.  However, the  discrete model operates with time steps $\Delta t = t_{k+1} - t_{k}$. While the time-continuous setting led to complex notation for evolving configurations, the time-discrete setting allows some simplifications in this regard. The above four processes are implemented in parallel over $[t_k,t_{k+1}]$. We note that the time discretization reflects a time-discontinuous process of growth and calcification.

\subsubsection*{Surface growth} We assume that at time $t_k$, the shell has been fully calcified: $\mathscr{C}_{t_k} = \Bvarphi(\Gamma_{t_k})$. In the time interval $[t_k, t_{k+1}]$, the leading surface is displaced due to mantle extension by $ v_1\Delta t\bs_1$ from $\omega_{t_{t_k}}$ (expressed as a deformed configuration) to a new reference surface $\omega_{0_{t_{k+1}}}$. This allows us to define a strip of the mantle in its reference configuration $\Omega^\text{m}_{0_{t_{k}}}$ bounded by the surface $\omega_{t_{t_k}}$ at its trailing edge and $\omega_{0_{t_{k+1}}}$ at its leading edge. See Fig. \ref{fig:fronts}.

\subsubsection*{Volume growth} Volume growth of the mantle is obtained by integrating Eqn (\ref{Eq:rateLg}). We exploit the exponential map:
\begin{equation}
    \bF^\text{g}_{t_{k+1}} = \bF^\text{g}_k\exp[\varepsilon_{2_k}\Delta t\bs_2\otimes\bs_2].
    \label{Eq:expFg}
\end{equation}
\noindent Since each increment of volume growth over a time step $\Delta t = t_{k+1} - t_k$ occurs relative to a new reference configuration, e.g., $\Omega^\text{m}_{0_{t_{k}}}$, we have $\bF^\text{g}_k = \bone$. 

The actual deformation gradient achieved is $\bF_{k+1}$, with the elastic component $\bF^\text{e}_{t_{k+1}} = \bF_{t_{k+1}}\bF_{t_{k+1}}^{\text{g}^{-1}}$ being governed by nonlinear elasticity. With a strain energy density function $\psi(\bF^\text{e})$ that satisfies frame invariance (so that $\psi(\bF^\text{e}) = \tilde{\psi}({\bF^\text{e}}^\text{T}\bF^\text{e})$), the first Piola-Kirchhoff stress tensor is 
\begin{equation}
    \bP = \partial\psi/\partial\bF^\text{e}.
    \label{Eq:P}
\end{equation}
It is governed by the quasistatic stress equilibrium equation imposed at time $t_{k+1}$:
\begin{equation}
    \text{Div}\,\bP_{t_{k+1}} = 0\;\text{in}\; \Omega^\text{m}_{0_{t_{k}}}.
    \label{Eq:goveq}
\end{equation}
In our computations we apply Dirichlet boundary conditions $\bu = \bzero$ on the trailing surface (boundary) $\omega_{t_{t_k}}$ of the mantle where it meets the rigid shell. A combination of fixed Dirichlet, $\bu = \bzero$, and traction-free Neumann conditions, $\bP\bN\big\vert_{t_{k+1}} = \bzero$ are applied on the remaining surfaces (boundaries) $\partial\Omega^\text{m}_{0_{t_{k+1}}}\backslash\omega_{t_{t_k}}$. This defines the morphoelastic growth problem for mapping the mantle strip from its reference configuration $\Omega^\text{m}_{0_{t_{k}}}$ to its deformed configuration $\Omega^\text{m}_{t_{t_{k}}}$.

\subsubsection*{Secretion}  Following morphoelastic volume growth of the mantle in the algorithmic step described above, a virtual step occurs, in which new material is secreted over the deformed configuration of the mantle $\Omega^\text{m}_{t_{t_{k}}}$.

\subsubsection*{Calcification}The final step of the algorithm is propagation of the calcification front so that $\mathscr{C}_{t_{k+1}} = \Bvarphi(\Gamma_{t_{k+1}})$. The mantle strip is calcified into its deformed configuration, $\Omega^\text{m}_{t_{t_{k}}}$.

\noindent\textbf{Remark 1}: The above algorithm is a manifestation of our observation that it is over the mantle that both surface growth and morpho-elastic volume growth occur. The actual formation of new shell material by secretion over the mantle, and the calcification of this material, follow once the current mantle configuration has been defined by surface and volume growth.

\noindent\textbf{Remark 2}: The steps presented above impose full calcification of the secreted material in deformed configuration $\Omega^\text{m}_{t_{t_{k}}}$. Consequently, the mantle in its reference configuration $\Omega^\text{m}_{0_{t_{k+1}}}$ attaches to the rigid material surface $\omega_{t_{t_{k+1}}}$. An alternate model with possible biological relevance is to assume that $\Omega^\text{m}_{{t}_{t_{k}}}$ has not been calcified, but remains elastic. Then the attachment of the mantle reference configuration $\Omega^\text{m}_{0_{t_{k+1}}}$ at time $t_{k+1}$ and its morphoelastic volume growth over $[t_{k+1},t_{k+2}]$ further deforms $\Omega^\text{m}_{t_{t_{k}}}$, also. Equation (\ref{Eq:goveq}) is then to be solved over $\Omega^\text{m}_{t_{t_{k}}}\cup\Omega^\text{m}_{0_{t_{k+1}}}$. Variants on this idea also are admissible, including complete calcification of a proper subset of $\Omega^\text{m}_{t_{t_{k}}}$ by time $t_{k+1}$, so that $\mathscr{C}_{t_{k+1}}$ does not coincide with $\Bvarphi(\Gamma_{0_{t_{k+1}}})$.

 \subsubsection*{Implementation}
\begin{figure}[!ht]
 \begin{center}
\psfrag{a}{\small{\shortstack[l]{First front: Reference \\configuration}}}
\psfrag{b}{\small{\shortstack[l]{First front: Deformed \\configuration}}}
\psfrag{e}{$\bF^\text{g},\bF$}
\psfrag{f}{$v_1\bs_1,\bF^\text{g},\bF$}
\psfrag{g}{$v_1\bs_1,\bF^\text{g},\bF$}
\psfrag{c}{\small{\shortstack[l]{Second front: Deformed \\configuration}}}
\psfrag{d}{\small{\shortstack[l]{Third front: Deformed \\configuration}}}
\psfrag{h}{\small{\shortstack[c]{Calcified\\$\bv^\text{c}\Delta t$}}}
\psfrag{i}{\small{\shortstack[c]{Calcified\\$2\bv^\text{c}\Delta t$}}}
 \includegraphics[width=0.8\textwidth]{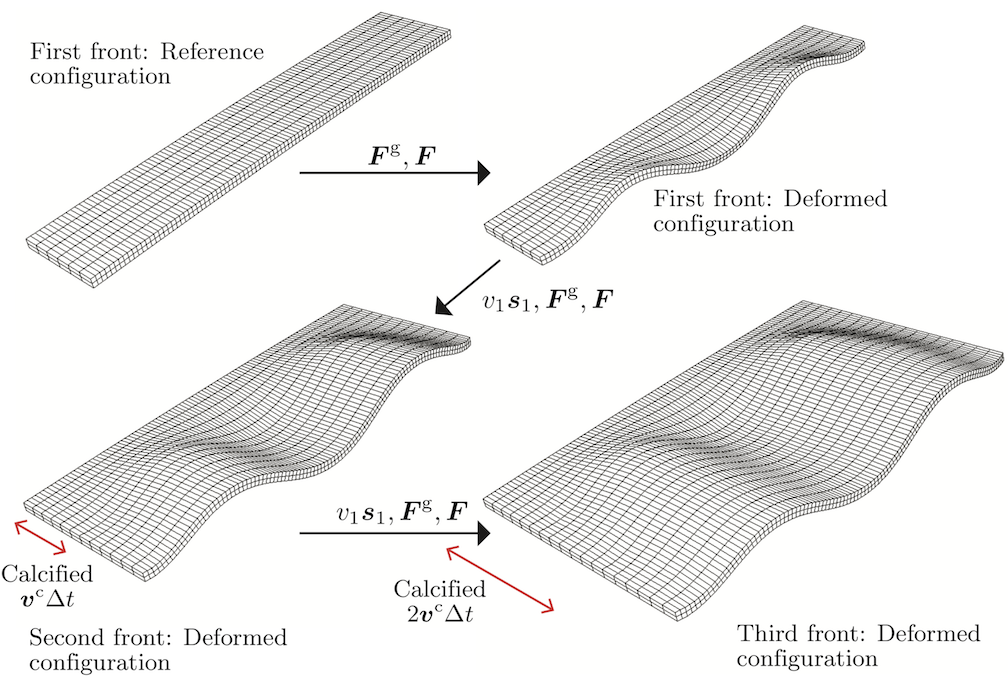}
 \caption{{\bf Space-time discretization in the finite element framework:} Evolution of the mollusk shell surface through surface growth and morphoelastic volume growth of mantle strips followed by their calcification. Also see S1 Movie for the time evolution of a representative molluskan shell surface through the accretive growth of 20 mantle strips.}
 \label{fig:mesh}
 \end{center}
 \end{figure}

The above formulation for surface and volume growth has been implemented in a finite element framework. An in-house C++ code based on the deal.II \cite{dealII90} open source finite element library is used to implement the model of surface and volume growth depicted in Figure ~\ref{fig:mesh}. Key highlights of this computational implementation are the ability to handle growing meshes (to model the growth of the reference configuration in a Lagrangian setting) and related dynamic updates to the solution data structures (global vectors and matrices). Simulations presented in this work use hexahedral elements with a linear/quadratic Lagrange basis, and one to four layers of elements through the thickness. The code base is publicly available as a GitHub repository \cite{github2019}.\\

The attachment of the mantle to $\omega_{t_{t_k}}$ and its extension up to $\omega_{0_{t_{k+1}}}$ is implemented by extending the finite element mesh by one or more rows of elements as shown in Fig \ref{fig:mesh}. This is followed by the growth law in Equation \eqref{Eq:expFg} subject to the constitutive law Equation \eqref{Eq:P} and the governing equations \eqref{Eq:goveq}. Following Remarks 1 and 2, secretion of shell material is a virtual step over the deformed mantle configuration $\Omega^\text{m}_{t_{t_{k}}}$. Calcification is imposed by turning $\Omega^\text{m}_{t_{t_{k}}}$ rigid for time $t \ge t_{k+1}$.  A number of examples are considered of mollusk shells displaying the shapes and marginal ornamentation that best demonstrate the interplay between surface growth, morphoelastic volume growth and the reference generating curve, $\Gamma_0$. In each case, the boundary condition is $\bu = \bzero$ on the trailing surface (boundary) $\omega_{t_{t_k}}$ of the mantle where it meets the rigid shell. The lateral surfaces (boundaries), $\partial\Omega^\text{m}_{0_{t_{k}}}\backslash\omega_{t_{t_k}}\backslash \omega_{0_{t_{k+1}}}$, are subject to Dirichlet conditions on either $\bu$, or its normal component, with traction-free Neumann conditions, $\bP\bN = \bzero$ on the remaining surfaces (boundaries) $\omega_{0_{t_{k+1}}}$. See Fig. \ref{fig:mesh} for an illustration of the evolving mantle and calcified shell configurations.

\section*{Results}

\begin{figure}[!ht]
 \begin{center} 
 \psfrag{a}{$\delta g$}
 \psfrag{b}{$\delta \bs$}
 \psfrag{c}{Calcified shell}
 \psfrag{d}{$\Gamma_0$, curvature $\kappa$}
 \psfrag{e}{}
 \psfrag{u}{$\bs_2$}
 \psfrag{v}{$\bs_1$}
 \psfrag{w}{$\bs_3$}
 \includegraphics[width=0.8\textwidth]{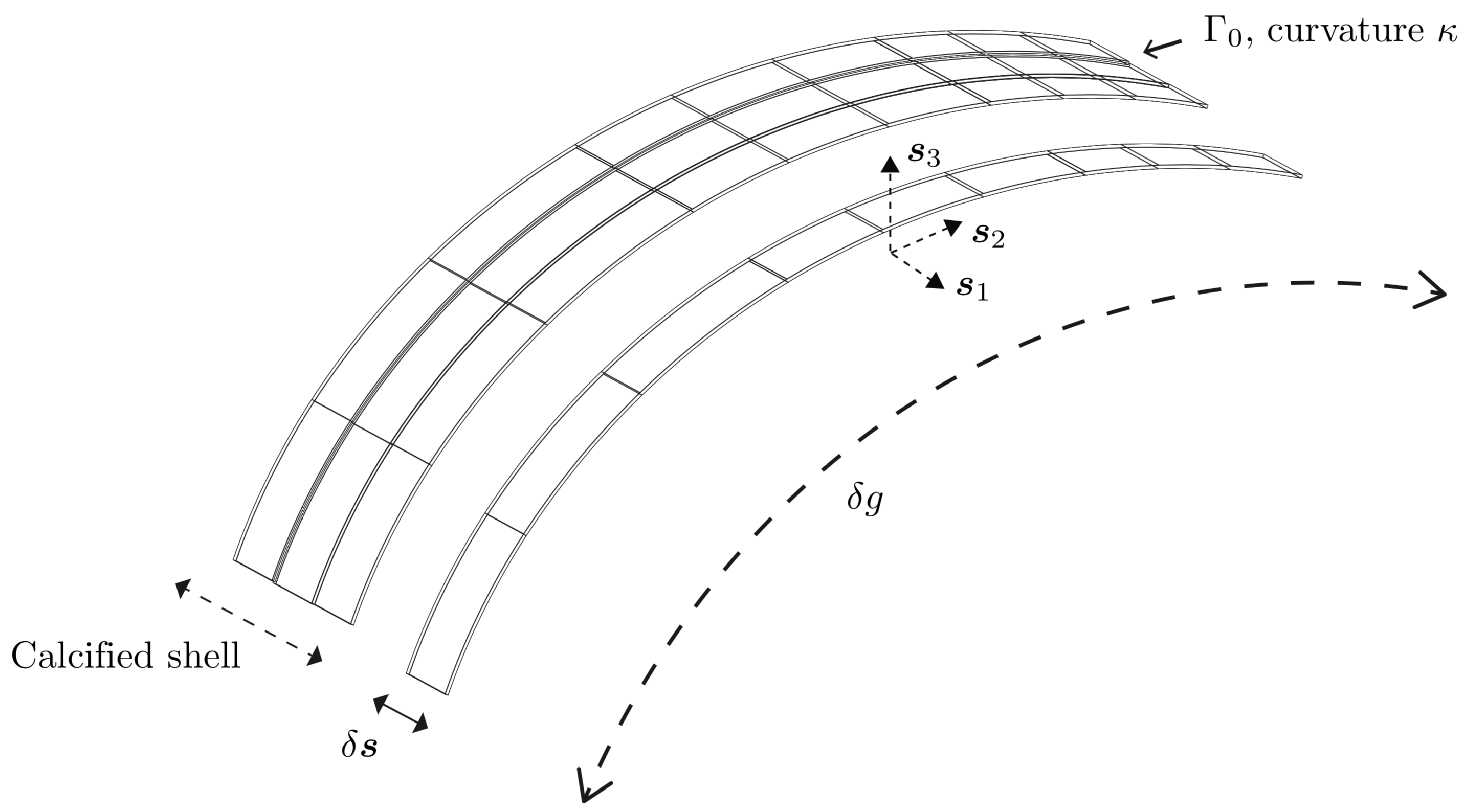}
 \caption{{\bf Parameters controlling the morphology of shell ornamentations:} The reference generating curve, $\Gamma_0$, with its curvature, $\kappa$ along $\bs_2$; the active mantle width, $\delta s$; volume growth strain increment, $\delta g$.}
 \label{fig:parameters}
 \end{center}
 \end{figure}

In the framework constructed above, there are three key parameters  determining the ornamentation pattern that develops as the shell grows:
\begin{enumerate}
   \item The \emph{active mantle width}, i.e. the amount of surface growth occurring in each time increment, given by $\delta s = \vert v_1\bs_1 - \bv^\text{c}\vert\Delta t$.
    \item The \emph{volumetric growth increment} over each time increment: $\delta g$, which is related to the growth strain rate $\dot{\bF}^\text{g} = \varepsilon_2\bF^\text{g}\bs_2 \otimes \bs_2$ by $\delta g = \varepsilon_2\Delta t$;
    \item The {\it initial curvature}, $\kappa$, of $\Gamma_0$.
\end{enumerate} 
These three governing parameters are illustrated in Fig.~\ref{fig:parameters}. Our objective is to explore the morphological space of patterns that results from variations in these parameters and explain them on a mechanistic basis while making connections to ornamentations observed on molluskan shells.

In the first instance, we study the morphologies that result from varying each parameter in isolation. In each case, we use our computational framework to impose either (1) a single, finite increment of surface growth manifested in a specified active mantle width, or (2) an increment of morphoelastic volume growth, or (3) observe the influence of the distribution of curvature along $\bs_2$ on the reference curve, $\Gamma_0$. The effects will be characterized by the mode number (the number of crests and valleys) along the lip of active mantle in the $\bs_2$ direction, the amplitudes and the locations of crests and valleys. In  a growing molluskan shell, these effects are coupled, and potentially dynamically changing through development. We therefore proceed next to analyze the coupled effects of variations in the above three parameters, as well as of spatially and temporally varying surface and volume growth.

\subsection*{Effect of parameters on morphology}
\label{sec:effectOfParameters}
\subsubsection*{Variation in surface growth via the active mantle width}
\label{sec:surfgrowth}
\begin{figure}[!ht]
 \begin{center}
 \psfrag{u}{$\bs_2$}
 \psfrag{v}{$\bs_1$}
 \psfrag{w}{$\bs_3$}
 \psfrag{a}{$\delta s = \delta s^\ast$}
 \psfrag{b}{$\delta s = 2\delta s^\ast$}
 \psfrag{c}{$\delta s = 4\delta s^\ast$}
 \psfrag{d}{$\delta s = 8\delta s^\ast$}
\includegraphics[width=0.9\textwidth]{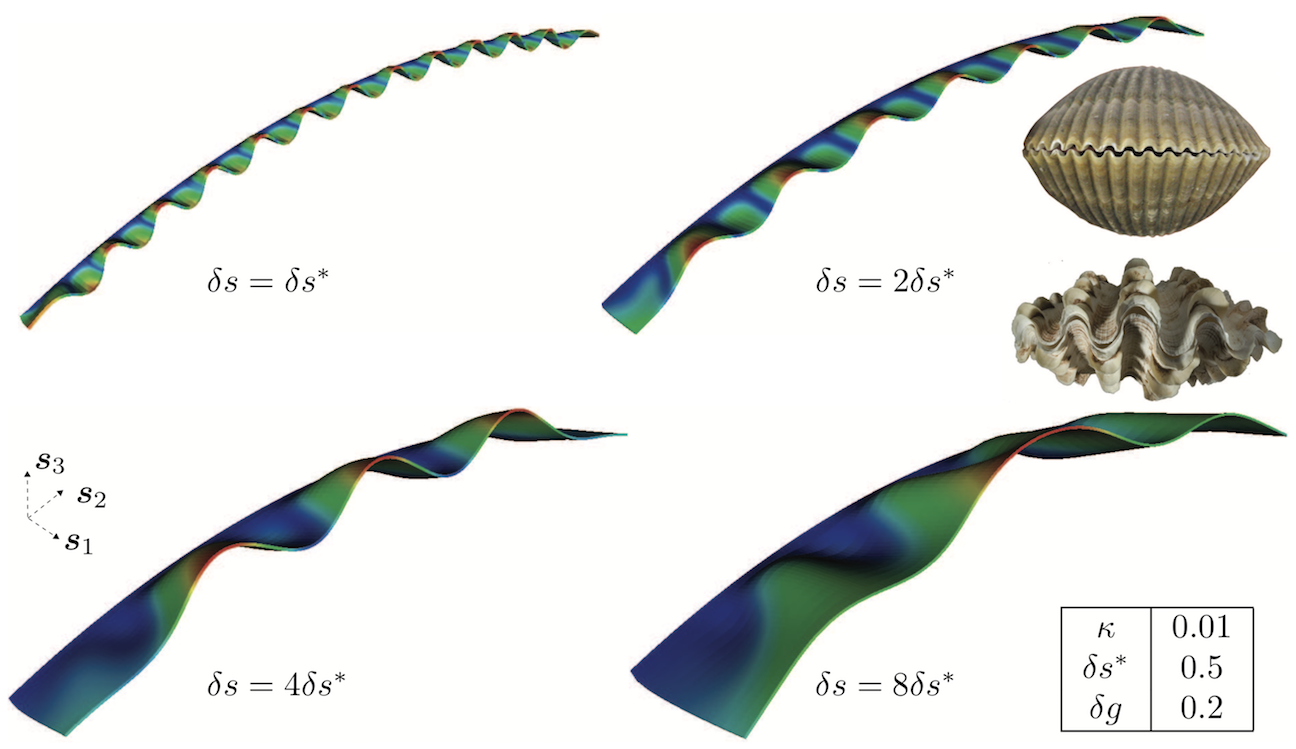}
 \caption{{\bf Effect of incremental active mantle widths on mantle deformation for fixed} $\delta g$ \textbf{and} $\kappa$: Increasing the active mantle width $\delta s = \vert v_1\bs_1 - \bv^\text{c}\vert\Delta t$ over $\Gamma_0$ leads to morphologies bearing a similarity with the ornamentations on \emph{\textit{Clinocardium nuttallii}} (upper inset) and \emph{\textit{Tridacna squamosa}} (lower inset). Dirichlet boundary conditions, $\bu = \bzero$, are applied on the trailing surface (boundary) and traction-free Neumann conditions, $\bP\bN = \bzero$, are applied on the remaining surfaces (boundaries). See Fig.~\ref{fig:fronts} for location of the trailing surface, front surface and the lateral surfaces. Also see S3 Movie for a morphology that is similar to the case $\delta s = \delta s^\ast$, and bears comparison to the ornamentation on members of the class \emph{Bivalvia}.}
 \label{fig:arcMode}
 \end{center}
 \end{figure}

The effect of varying surface growth by the active mantle width is studied for fixed volume growth rate and reference curvature. With our computational formulation, we solve for the resulting shape after a single surface growth increment of active mantle. We start from a reference generating curve, $\Gamma_0$, which is an arc of a circle with curvature $\kappa=0.01$. Fig.~\ref{fig:arcMode} shows the resulting morphologies when the active mantle width $\delta s = \vert v_1\bs_1 - \bv^\text{c}\vert\Delta t$ is varied by up to eightfold. We see that an increase in $\delta s$ leads to a decrease in mode number, which can be understood as follows: increasing $\delta s$ places the free edge of the active mantle strip, $\omega_{0_{t_{k+1}}}$, further from the rigid boundary $\omega_{t_{t_k}}$ ($k = 1,2,\dots$) in the $\bs_1$-direction, thus decreasing its structural stiffness to bending. The excess material (along $\bs_2$) created by the increment in volume growth strain $\delta g$ can therefore be accommodated by an increased deformation in the $\bs_3$ direction, without paying a large strain energy penalty. As a result, each increment in $\delta s$ induces fewer wave crests/valleys, with progressively larger wavelength and amplitude. Fig. 7 also presents a comparison with the ornamentations on \emph{\textit{Clinocardium nuttallii}} and \emph{\textit{Tridacna squamosa}}, the active mantle width being much larger in the second species (giant clam), and both species differing in amplitude and wavelength of the antimarginal ribs in a manner that is consistent with the model predictions.

Here, it is instructive to compare to an analogous reduced order model: a growing one-dimensional rod on an elastic foundation. In this model, an elastic rod that has an excess of length due to axial growth is connected elastically to a rigid support: a curve representing the calcified shell edge. The support provides an external force that resists displacement of the rod away from the foundation (i.e. displacement in the $\bs_3$ direction in our framework). This system has been formulated in detail by some of the authors \cite{moulton2012morphoelastic} and forms the basis of previous mechanical descriptions of shell morphogenesis \cite{erlich2018mechanical,chirat2013mechanical}. In the linearized system, with the rod and foundation extending along the $x$-axis, the deformed rod has shape $y(x)$ satisfying \cite{moulton2012morphoelastic}

\begin{equation}\label{equation:Rod}
y''''(x) + (\gamma-1) y''(x) + k\gamma y(x) = 0,
\end{equation}
 
Here $\gamma>1$ describes the axial growth, analogous to $\delta g$ in the computational framework. The parameter $k$ is proportional to the stiffness of the elastic foundation, and therefore models the stiffness of the active mantle strip to deflections of the mantle margin.
Considering for simplicity an infinite rod and seeking a solution of the form $y\sim \exp(2\pi i n x)$, the preferred bifurcation mode corresponds to the smallest value of $\gamma^*>1$ for which \eqref{equation:Rod} has a solution; this is found to be $\gamma^*=1+2k+2(k+k^2)^{1/2}$, from which we obtain that the mode number at buckling satisfies $n=\sqrt{\gamma^*-1}/(2\sqrt{2}\pi)$. From this we can extract the scaling relationship $n\sim \sqrt{k}$ for large $k$. 

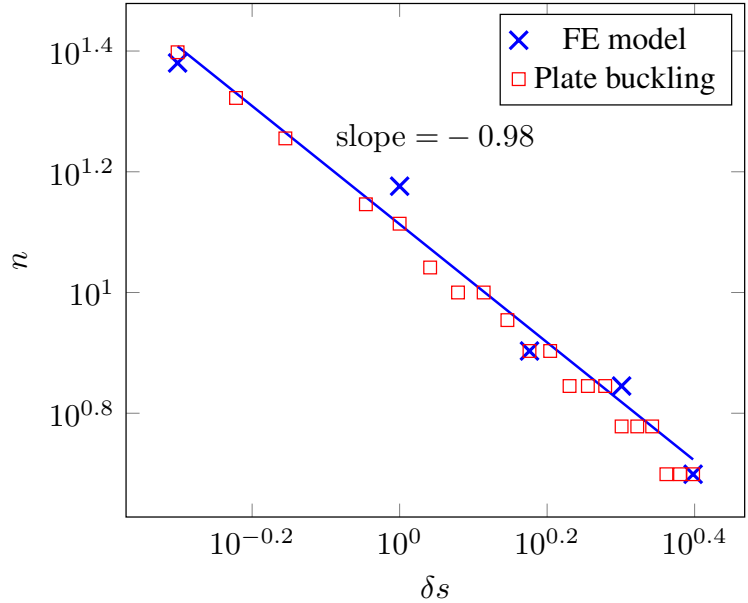
\begin{figure}[!ht] 
  \centering    
    \begin{tikzpicture}[scale=1.2]
      \begin{loglogaxis}[minor tick num=1,xlabel={$\delta s$},ylabel={$n$}]
    \addplot [only marks,color=blue,mark = x,mark size=4pt, line width=1pt] table {\tableMantleWidth};
    \addplot [only marks,color=red,mark = square,mark size=2pt] table {\tableMantleWidthPlate};
    \legend{FE model, Plate buckling}
    \addplot [thick, blue, forget plot] table[
    y={create col/linear regression={y=Y}}] {\tableMantleWidth}
    node [pos=0.5, above=28pt, color=black] {$\mathrm{slope=} \pgfmathprintnumber[precision=2, fixed zerofill]
      {\pgfplotstableregressiona}$};
       \end{loglogaxis}
    \end{tikzpicture}
  \caption{\textbf{Scaling study of mode number $n$ \emph{versus} the incremental active mantle width $\delta s$}. Shown is the dependence of the mode number of the deformed mantle strip on the incremental active mantle width, obtained from the finite element model (FE) and the buckling analysis of a plate.}
  \label{fig:scaling2}
\end{figure}

Based on the intuitive argument above, we would posit an inverse relationship between $k$ and $\delta s$, e.g. $k\sim\delta s^\alpha$ with $\alpha<0$, i.e. an increase in strip width acts to decrease the effective foundation stiffness, leading to a decrease in bifurcation mode. To further explore this relationship, we extract the dependence of mode number $n$ on $\delta s$ from the simulations presented in Fig.~\ref{fig:arcMode}, and plot the comparison in Fig.~\ref{fig:scaling2}. Because of the highly nonlinear, post-bifurcation states of deformation, $n$ was defined as the number of waves, following crests or troughs, and ignoring the dependence of amplitude and wavelength on the coordinate in the $\bs_2$-direction. To validate the computational model, we also include the critical mode as calculated from a buckling analysis of a plate (see S1 Text).
From the slope of this log-log plot, we get $n\sim \delta s^{-1}$, and so $k \sim \delta s^{-2}$. In principle, one could use this map to more systematically parameterize the foundation in the 1D morphoelastic rod framework. Computing the morphology of the shell edge as a 1D (geometrically nonlinear) structure has the advantage of decreased computation time, though with potential inaccuracies due to loss of detail. A systematic means of determining $k$ provides a very useful step in alleviating this, though it remains an interesting question how far into the post-buckling regime the relation $k \sim \delta s^{-2}$ holds.

\pgfplotstableread{
X Y
0.5 24
1.0 15
1.5 8
2.0 7
2.5 5
}\tableMantleWidth

\pgfplotstableread{
X Y
0.5 25
0.6 21
0.7 18 
0.9 14
1.0 13
1.1 11
1.2 10
1.3 10
1.4 9
1.5 8
1.6 8
1.7 7
1.8 7
1.9 7
2.0 6
2.1 6
2.2 6
2.3 5
2.4 5
2.5 5
}\tableMantleWidthPlate

\subsubsection*{Morphoelastic volume growth rate}
\label{sec:volgrowth}
 In our model, the morphoelastic volume growth of the soft mantle is the origin of the pattern of bifurcation: This mechanism generates an excess of length in the active mantle strip relative to the rigid shell edge, whose distribution is given by the variation of $\delta g$ along $\bs_2$. A compressive stress is thus induced in the mantle. As is well-understood within the theory of finite strain elasticity, when this in-surface stress exceeds a critical threshold, a bifurcation occurs and the mantle relaxes into a lower energy configuration by deflecting out of the mid-surface in the local $\bs_3$ direction. 
 
\begin{figure}[!ht]
 \begin{center}
  \psfrag{a}{\begin{tabular}{@{}l@{}} $\delta g = \delta g^\ast$ \\  $n = 3$ \end{tabular}}
  \psfrag{b}{\begin{tabular}{@{}l@{}} $\delta g = 2 \delta g^\ast$ \\  $n = 7$ \end{tabular}}
  \psfrag{c}{\begin{tabular}{@{}l@{}} $\delta g = 3 \delta g^\ast$ \\  $n = 7$ \end{tabular}}
  \psfrag{d}{\begin{tabular}{@{}l@{}} $\delta g = 4 \delta g^\ast$ \\  $n = 7$ \end{tabular}}
 \psfrag{u}{$\bs_2$}
 \psfrag{v}{$\bs_1$}
 \psfrag{w}{$\bs_3$}
 \includegraphics[width=0.9\textwidth]{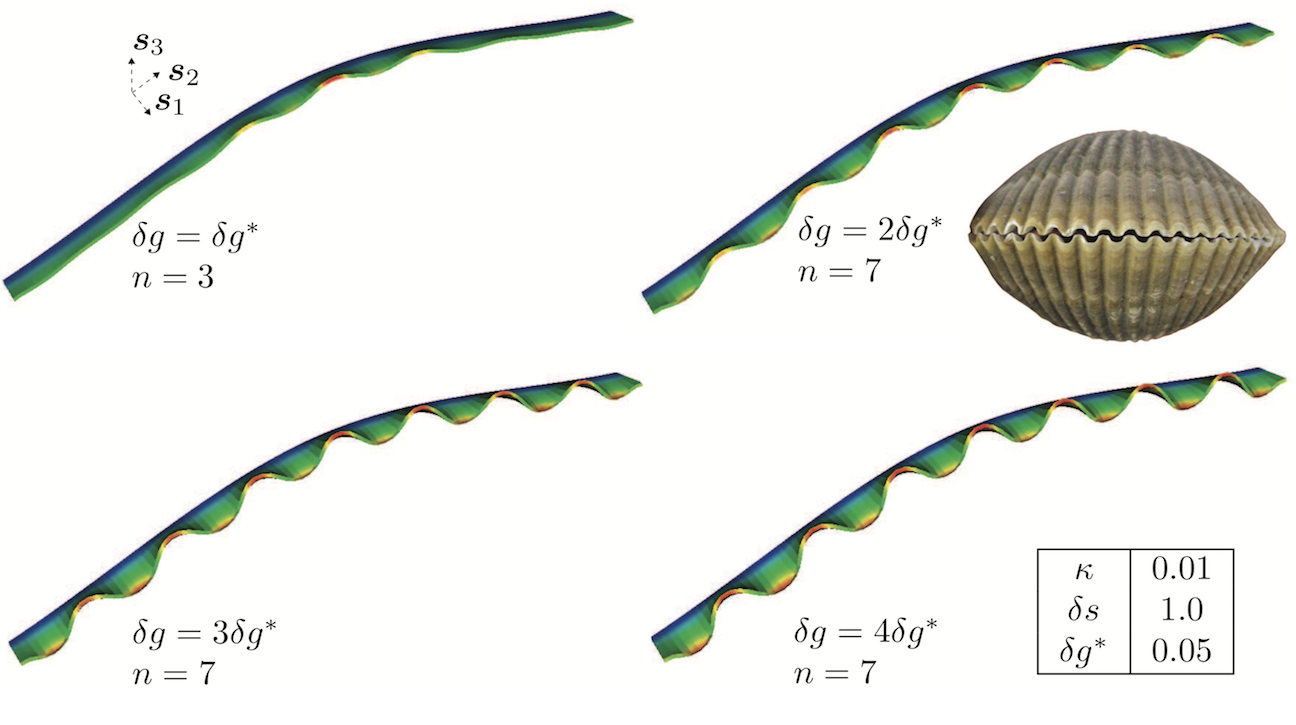}
 \vspace*{1mm}
 \caption{{\bf Effect of volume  growth strain increment}, $\delta g = \varepsilon_2\Delta t$ on the amplitude and mode numbers of the deformed mantle. The observed high mode number morphologies are similar to the ornamentations observed in bivalves like \emph{\textit{Clinocardium nuttallii}} (inset).
 Dirichlet boundary conditions, $\bu = \bzero$, are applied on the trailing surface (boundary) and traction-free Neumann conditions, $\bP\bN = \bzero$, are applied on the remaining surfaces (boundaries). See Fig.~\ref{fig:fronts} for location of the trailing surface, front surface and the lateral surfaces.}
 \label{fig:growthRate}
 \end{center}
 \end{figure}

 Fig.~\ref{fig:growthRate} illustrates the effect of varying the incremental growth strain $\delta g$ by up to a factor of $4\times$ while holding the incremental active mantle width fixed at $\delta s=1.0$. The reference generating curve $\Gamma_0$ has curvature of $\kappa = 0.01$ in the middle of its extent in the $\bs_2$ direction, decreasing toward zero near the ends. Note that this induces a length scale: the radius of curvature $r = 100$, relative to which $\delta s = 1$.

In a time-continuous process, there would exist a critical time, $t_\text{cr}$, and corresponding amount of morphoelastic volume growth strain, $ \delta g_\text{cr} = \int_0^{t_\text{cr}} \varepsilon_2(t) \mathrm{d}t$ for which the compressive stress crosses the critical threshold and the corresponding bifurcation mode appears. Modes with greater numbers of waves have increased energy. Then, as volume growth continues beyond  $\delta g_\text{cr}$ the compressive mantle stress and amplitude of the first observed bifurcation mode both increase, until the stress exceeds a second critical threshold corresponding to another bifurcation and a higher mode appears. This effect is demonstrated in Fig~\ref{fig:growthRate}, where initially the compressive stress (corresponding to $\delta g = \delta g^\ast$) initiates bifurcation into a mode with three discernible crests, $n = 3$. A growth strain increment to $\delta g = 2\delta g^\ast$ gives a compressive mantle stress that exceeds a higher critical threshold, and is accommodated by an increase in the mode number to $n = 7$. Increasing the growth to $\delta g = 3\delta g^\ast, 4 \delta g^\ast$ only increases the amplitude of the seventh mode, with no further bifurcations.

\noindent\textbf{Remark 3}: A careful examination of the bifurcated shape with mode number $n = 3$ for $\delta g = \delta g^\ast$ reveals a deformation with amplitude that is highest at the midpoint of the arc of $\Gamma_{0}$, where the curvature $\kappa = 0.01$, decreases in the two immediate neighbor crests, and decays by more than an order of magnitude toward the lateral edges, where $\kappa \to \infty$. Inclusion of all crests regardless of amplitude would raise the inferred mode number to $n = 7$ even for this first volume growth increment. We understand this as a cascade in which the lowest mode to appear ($n =3$) is localized to the higher curvature region. At $\delta g = \delta g^\ast$, there is a super-position, with the $n = 5$ and $n = 7$ modes also present, but at lower amplitude. The coincidence of crests for modes $n = 3, 5, 7$ in the high curvature region leads to the higher amplitudes there. At the very next increment to $\delta g = 2\delta g^\ast$ the strain energy settles into the seventh mode, whose prominence is magnified. In contrast, for $\kappa = 0.01$, but uniform as in Fig. \ref{fig:arcMode}, a single mode is observed, whose amplitude is uniform provided it does not merge into the lateral boundary. Also consider Fig. \ref{fig:baseGeometry}a with $\kappa \to 0$ and uniform, where the amplitude remains uniform. The localization of mode shapes is a consequence of curvature, which we examine in greater detail in \red{the next subsection on Curvature}.

\subsubsection*{Curvature}
\label{sec:curvature}
\begin{figure}[!ht]
 \begin{center}
 \psfrag{u}{$\bs_2$}
 \psfrag{v}{$\bs_1$}
 \psfrag{w}{$\bs_3$}
 \includegraphics[width=0.9\textwidth]{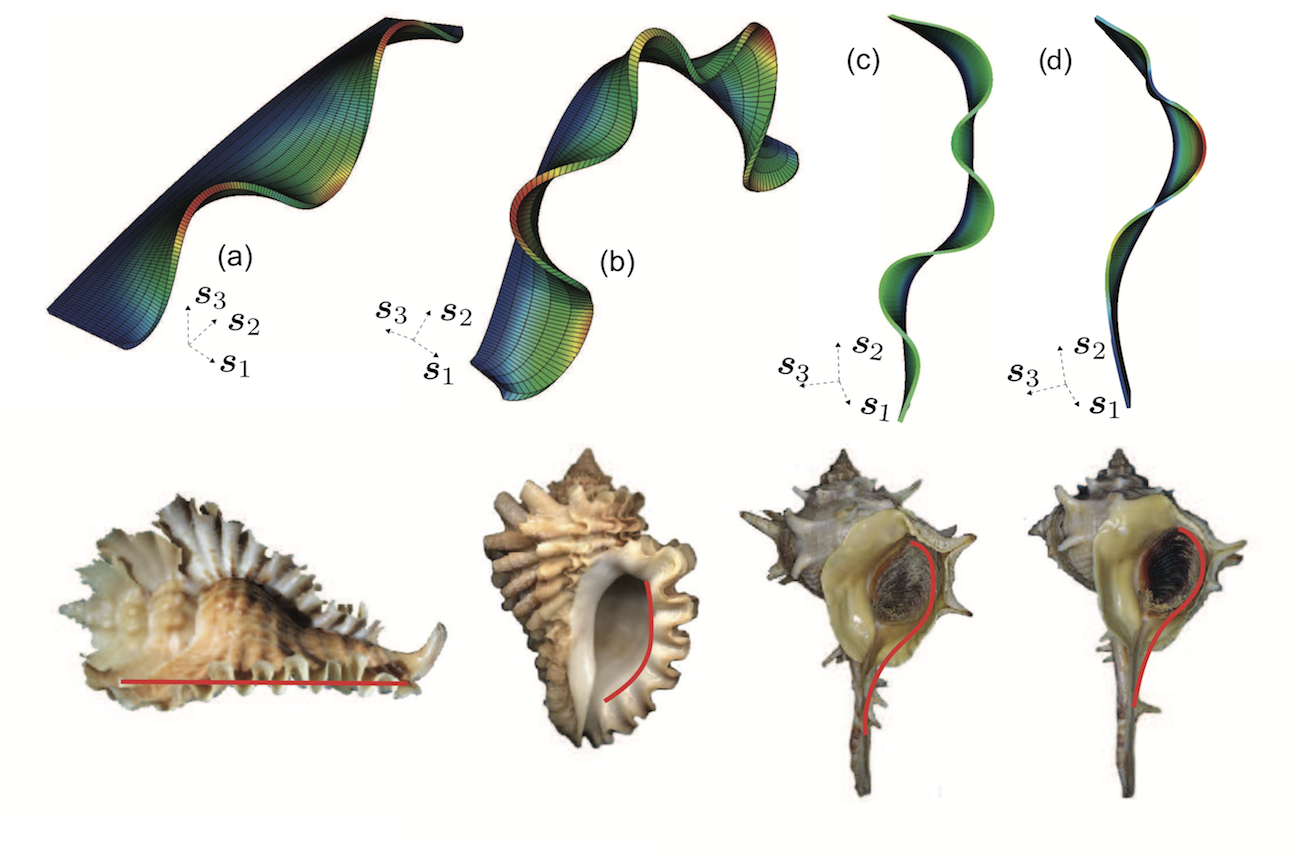}
 \caption{{\bf The influence of the geometry of reference curves on antimarginal ornamentation.} For fixed active mantle width, the amplitudes of crests in the deformed configurations are  magnified if they are compatible, and attenuated if they are incompatible, respectively, with the reference generating curve. These reference curves bear similarity to the shape of the mantle surface (highlighted in red) found in (a) \emph{Pterynotus phyllopterus}, (b) \emph{Nucella freycineti}, (c) \emph{Normal Bolinus brandaris} and (d) \emph{Abnormal Bolinus brandaris} (insets). Dirichlet boundary conditions, $\bu = \bzero$, are applied on the trailing surface (boundary) and the lateral surfaces (boundaries) of the mantle, which are perpendicular to $\bs_2$, and traction-free Neumann conditions, $\bP\bN = \bzero$, are applied on the front surface (boundary). The underlying spatial discretization (mesh) is also shown on the model geometries. Also see S2 Movie-S5 Movie for the evolution of mantle deformation and accretive growth over planar, arc, and closed circular geometries of the reference curves.}
 \label{fig:baseGeometry}
 \end{center}
 \end{figure}

Taking a cue from the localization of modes in high $\kappa$ regions of the reference generating curve, $\Gamma_0$, we next consider this effect in greater detail. We consider three geometries for $\Gamma_0$: a line with curvature $\kappa \to \infty$, a curve with $\kappa$ having almost uniform sign, and a second curve with $\kappa$ changing sign along the arc. The result appears in Fig~\ref{fig:baseGeometry}. For all three cases in the figure, the Dirichlet boundary conditions are $\bu = \bzero$ on the trailing surface (boundary), $\omega_{0_{t_0}}$, and the lateral surfaces (boundaries) of the mantle, which are perpendicular to $\bs_2$.

We find that the deformed configuration of the mantle $\Omega^\text{m}_{t_{t_0}}$ is biased toward developing curvature of the same sign as the reference curvature, $\kappa$, along $\bs_2$.  With $\Gamma_0$ a straight line in Fig~\ref{fig:baseGeometry}a, the mantle deforms upward and downward equally, i.e. curvatures of both signs are seen in $\Omega^\text{m}_{t_{t_0}}$, and the crests/troughs have the same amplitude. The (mostly) single-signed $\kappa$ of Fig~\ref{fig:baseGeometry}b promotes like-signed crests and suppresses oppositely signed ones. This is also apparent in Fig~\ref{fig:baseGeometry}d (and to a lesser degree in Fig~\ref{fig:baseGeometry}c), which has regions where $\kappa$ takes on positive and negative signs. Thus, we see the influence of geometry in inducing compliance to mantle deformation by forming crests  that are compatible, and resistance to forming crests that are incompatible with the reference curvature, respectively. This pattern of deformation is consistent with the greater amplitude of the central, compatible crest in Fig. \ref{fig:growthRate} with $\delta g = \delta g^\ast$.

This result has an intriguing relevance for mollusk shell ornamentation: the reference shape of the shell on which ornamentation appears, modelled here by $\Gamma_0$, is generally convex and surrounds the mollusk body. It would be disadvantageous for the ornamentation to appear inward, as this would intrude on the mollusk's living space. A natural question then is whether the mollusk must execute a complex developmental process to ensure that the ornamentation is built in the outward direction. The results here suggest, rather, that the geometry and growth mechanisms naturally conspire to bias the pattern outward.

\begin{figure}[!ht]
 \begin{center}
 \psfrag{a}{\small Mesh}
 \psfrag{b}{\small $\delta g=0$}
 \psfrag{c}{\small $\delta g=\delta g^\ast$}
 \psfrag{d}{\small $\delta g=2\delta g^\ast$}
 \psfrag{e}{\small $\delta g=4\delta g^\ast$}
 \psfrag{f}{\small $\delta g=6\delta g^\ast$}
 \psfrag{g}{\small $\delta g=8\delta g^\ast$}
 \psfrag{u}{\small $\bs_2$}
 \psfrag{v}{\small $\bs_1$}
 \psfrag{w}{\small $\bs_3$}
 \psfrag{x}{\small $\delta s$}
 \includegraphics[width=0.9\textwidth]{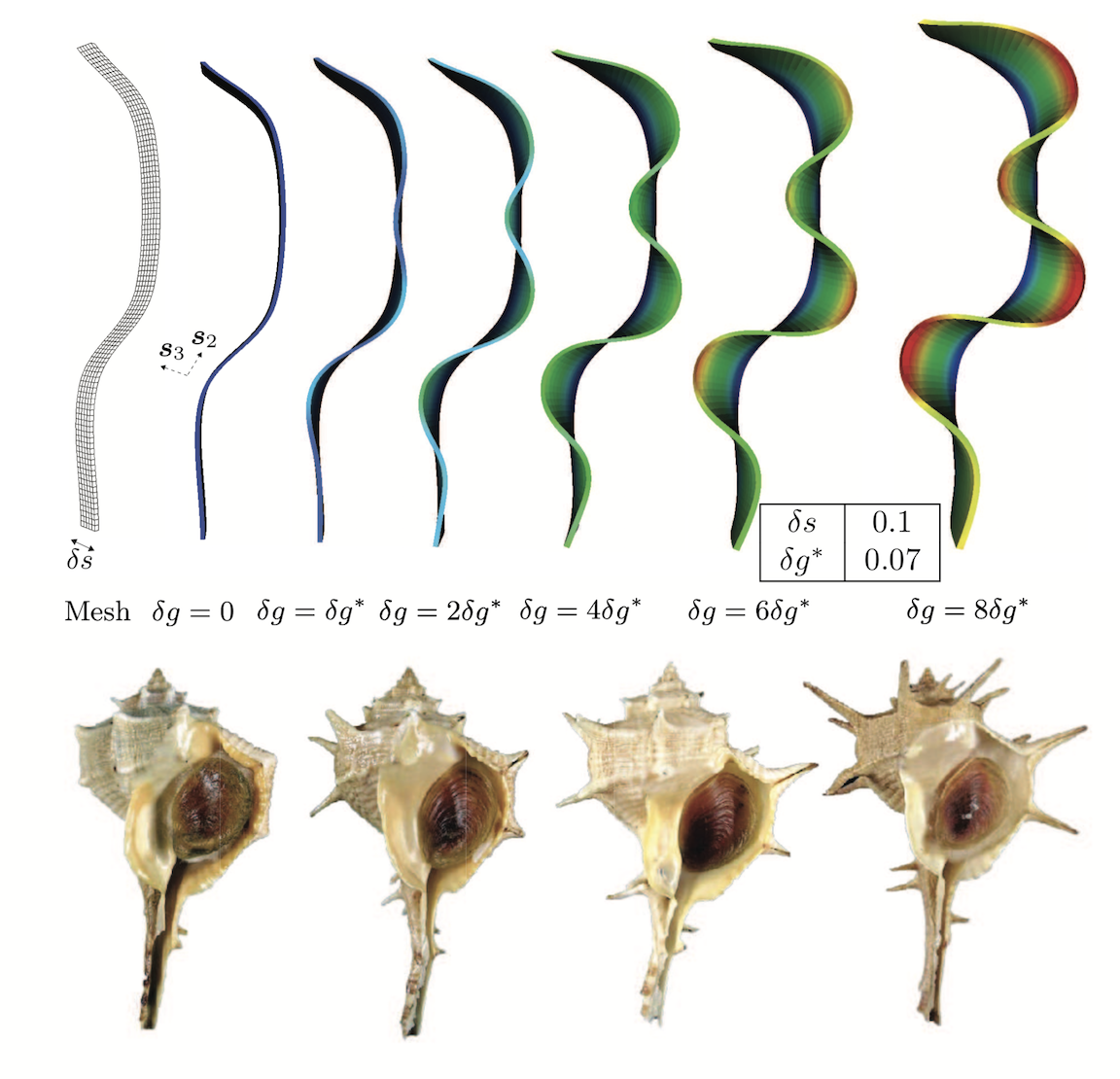} 
 \caption{{\bf Progression of curvature-compatible ornamentation with volume growth strain increments}, $\delta g$. The deformed mantles show marginally preferred localization around points of high curvature and thereafter the amplitude increases with volume growth strain increments. Some of these shapes with different amplitudes can be observed in the shells of the species \emph{Bolinus brandaris} (bottom row of inset images). Dirichlet boundary conditions, $\bu = \bzero$, are applied on the trailing surface (boundary) and traction-free Neumann conditions, $\bP\bN = \bzero$, are applied on the remaining surfaces (boundaries).}
 \label{fig:splineModes}
 \end{center}
 \end{figure}

\subsection*{More complex patterning}
The influences of the surface and volume growth rates, and of the geometry via reference curvature, have been established. We now consider the combination of these effects, and their temporal and spatial variations, in two mechanisms that lead to more complex ornamentations.
 
\subsubsection*{Progression of ornamentation with volume growth}
\begin{figure}[!ht]
 \begin{center}
  \psfrag{a}{\small Mild $\kappa$ singularities}
  \psfrag{b}{\small Strong $\kappa$ singularities}
  \psfrag{c}{\small Smooth curve}
 \psfrag{u}{$\bs_2$}
 \psfrag{v}{$\bs_1$}
 \psfrag{w}{$\bs_3$}
 \includegraphics[width=0.9\textwidth]{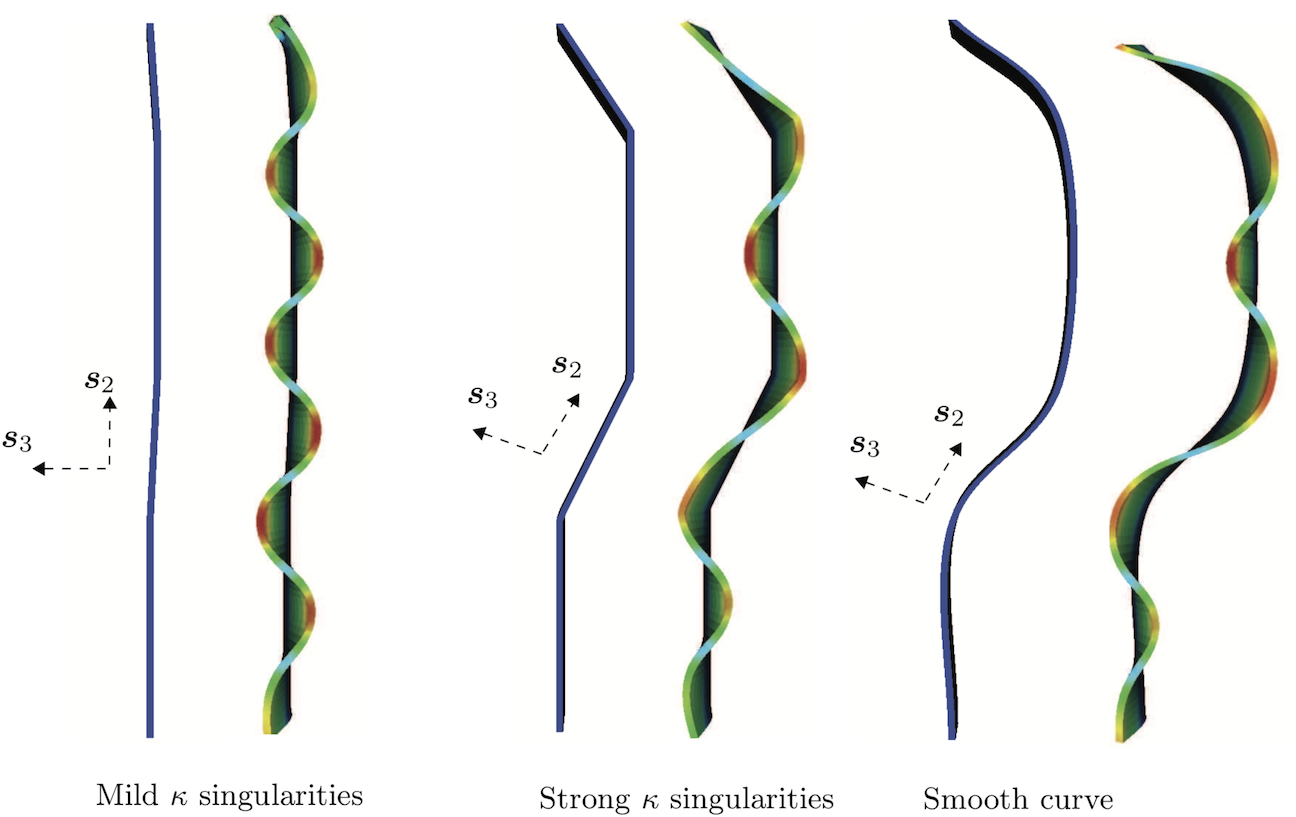} 
 \caption{{\bf Influence of reference curvature singularities, and of smooth curvatures}. Mild reference curvature singularities leave virtually no visible trace on mantle deformation following volume growth. However, strong reference curvature singularities promote compatible crests, and remain visible as mild singularities in the deformed mantle lip. Smoothly varying reference curvature also replicates the trend of favoring compatible crests. Dirichlet boundary conditions, $\bu = \bzero$, are applied on the trailing surface (boundary) and traction-free Neumann conditions, $\bP\bN = \bzero$, are applied on the remaining surfaces (boundaries).}
 \label{fig:curvatureModes}
 \end{center}
 \end{figure}

As demonstrated above, geometry exerts its influence by magnifying the  amplitudes of crests in mantle deformation that are compatible, and attenuating those crests that are incompatible, respectively, with reference curvature. It is of interest to study the progression of these crests with continued volume growth. With this aim, we consider a shell edge with the geometry of Fig~\ref{fig:baseGeometry}(c), having varying curvature, and impose volume growth strain increments ranging from $\delta g = 0.0$ to $\delta g = 0.56$. Several of these mantle deformations are shown in Fig~\ref{fig:splineModes}. The trend observed in Fig~\ref{fig:baseGeometry}---of magnification and attenuation of mantle deformation that is respectively compatible and incompatible with the reference curvature---continues. Favored crests display progressive magnification of amplitudes with continued growth. Also shown are \emph{Bolinus brandaris} shells with progressively increasing amplitudes of crests corresponding to the mantle deformations in our computations. \red{The pronounced localization into spines has been explained by some of the authors of this communication via the added mechanism of spatially varying material properties \cite{moulton2012morphoelastic}}. Fig~\ref{fig:curvatureModes} examines the smoothness of geometry. Mild reference curvature singularities leave virtually no visible trace on mantle deformation following volume growth. However, strong reference curvature singularities promote compatible crests, and remain visible as mild singularities in the deformed mantle lip. Smoothly varying reference curvature also replicates the trend of favoring compatible crests.
 
\subsubsection*{Hierarchical ornamentation by temporal variation of growth rates}
\begin{figure}[!ht]
\begin{center}
\psfrag{a}{$\delta g_0 = \delta g^\ast$ in $(t_0,t_1]$}
\psfrag{b}{\begin{tabular}{@{}l@{}} \small Surface growth of \\ second layer ($0.5\delta s^\ast$) \end{tabular}}
\psfrag{c}{$\delta g_1 = \delta g^\ast$ in $(t_1,t_2]$}
\psfrag{d}{$\delta g_2 = 2\delta g^\ast$ in $(t_2,t_3]$}
\psfrag{x}{$\delta s^\ast$}
\psfrag{y}{$1.0$}
\psfrag{z}{$1.5\delta s^\ast$}
\psfrag{e}{}
\psfrag{u}{$\bs_2$}
\psfrag{v}{$\bs_1$}
\psfrag{w}{$\bs_3$}
\psfrag{k}{$\Omega^\text{m}_{0_{t_0}}$}
\psfrag{l}{$\Omega^\text{m}_{t_{t_0}}$}
\psfrag{m}{$\Omega^\text{m}_{t_{t_0}}$}
\psfrag{n}{$\Omega^\text{m}_{0_{t_1}}$}
\psfrag{o}{$\Omega^\text{m}_{t_{t_0}}$}
\psfrag{p}{$\Omega^\text{m}_{t_{t_1}}$}
\psfrag{q}{$\Omega^\text{m}_{t_{t_0}}$}
\psfrag{r}{$\Omega^\text{m}_{t_{t_2}}$}
\psfrag{i}{}
\includegraphics[width=0.9\textwidth]{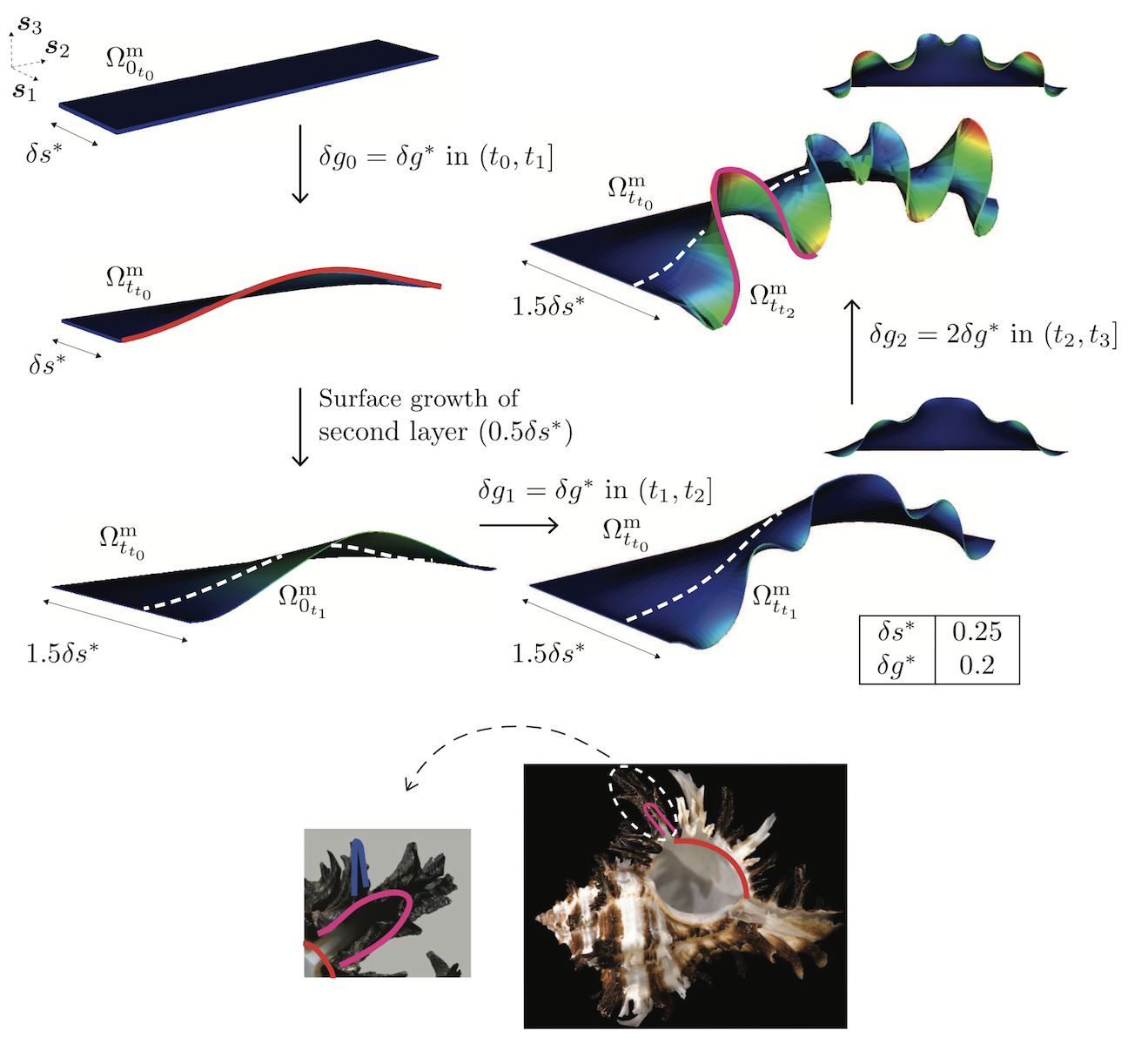}
\caption{\textbf{Hierarchical ornamentation arising from temporally varying surface growth, $\delta s$ volume growth strains, $\delta g$} $\varepsilon_2$. In the \emph{Hexaplex cichoreum} image shown in the inset, three levels of ornamentation hierarchy are shown: primary (red) as a low mode bifurcation from a flat surface, secondary ornamentation as a second mode bifurcation (magenta) and tertiary ornamentation mode as a third mode (blue). The corresponding first, second and third modes are traced on the mantle edge of the computational model. The dotted white line indicates the location of the fixed calcification front between ($t_1,t3$]. Dirichlet boundary conditions, $\bu = \bzero$, are applied on the trailing surface (boundary) and the lateral surfaces (boundaries) of the mantle, which are perpendicular to $\bs_2$, and traction-free Neumann conditions, $\bP\bN = \bzero$, are applied on the front surface (boundary). Inset image of \emph{Hexaplex cichoreum} modified from source \cite{HexaplexRef1}. Original images licensed under the Creative Commons Attribution-Share Alike License.} 
\label{fig:hierarchical}
\end{center}
\end{figure}

As a second approach to complex patterning we study the effect of multiple generations of surface and volume growth. Since the examples  \red{presented in the earlier sections} have already considered a range of either surface or volume growth rates varying individually, we now consider the effect of combining these growth rates while also allowing them to vary in time. Our aim in this section is to identify a mechanism that could explain the secondary and tertiary crests and valleys that are visible along the shell edge in species such as \emph{Hexaplex cichoreum}. We recognize these as the potential remnants of increasing mode number during shell growth. Noting that the decrease in mode number with increasing active mantle width, \red{as shown earlier,} implies an increase in mode number with decreasing active mantle width, and recalling the magnification of higher modes with increasing volume growth strain rates, we consider the following protocol of surface and volume growth rates in Fig~\ref{fig:hierarchical}: initial surface growth lays down a mantle of width $\delta s_0 = \delta s^\ast$, forming a reference configuration $\Omega^\text{m}_{0_{t_0}}$, which, under morphoelastic volume growth over $(t_0,t_1]$ deforms into $\Omega^\text{m}_{t_{t_0}}$. Upon calcification, the curved mantle edge $\omega_{t_{t_1}}$ provides the reference curvature for subsequent growth. The second reference configuration $\Omega^\text{m}_{0_{t_1}}$ laid down by surface growth has only half the initial mantle width: $\delta s_1 = 0.5\delta s^\ast$. However, with the same morphoelastic volume growth $\delta g_1 = \delta g^\ast$ over $(t_1,t_2]$, a mode of  higher mode number ($n = 3$) develops: secondary ornamentation is achieved on configuration $\Omega^\text{m}_{t_{t_1}}$. No further surface growth occurs, but  the volume growth undergoes another increment of $\delta g_2 = 2\delta g^\ast$ over $(t_2,t_3]$. Another bifurcation into a higher mode, $n = 4$, is seen and further detail of secondary ornamentation is visible on the configuration $\Omega^\text{m}_{t_{t_2}}$. In the interval $(t1,t_3]$, the calcification front is stationary at a distance of $\delta s^\ast$ along the nominal $\bs_1$ direction, as indicated by the dotted white line in Fig~\ref{fig:hierarchical}.

The resulting shell morphology thus has a hierarchical structure to its ornamentation, with higher modes appearing over configurations that initially have lower modes. Such features are present in the ornamentations of a number of muricid species including \emph{Hexaplex cichoreum} and \emph{Hexapelx duplex}. Indeed, as shown in the \emph{H. cichoreum} shell in Fig \ref{fig:hierarchical}, a tertiary bifurcation mode also appears, with some shells even showing quaternary modes in a fractal-like structure. Such morphological features are within reach of our model in principle, although resolving details beyond secondary modes becomes challenging due to the computational complexity associated with ensuring curvature continuity in the $\bs_1$ direction with accumulation of high curvature crests. The combined modulation of surface growth rate (active mantle width), volume growth rate and curvature presents a simple mechanical basis for the morphogenesis of hierarchical ornamentation in seashells, which has not previously been described.

\subsubsection*{Ornamentation with negative Gaussian curvature due to spatial variation in growth rate}
An examination of the mantle deformation in Fig. \ref{fig:arcMode} and Figs \ref{fig:growthRate}-\ref{fig:hierarchical} shows that the majority of crests and valleys form with positive Gaussian curvature. One exception is Fig. \ref{fig:baseGeometry}b. This case is explained by the strain energy due to high curvature, $\kappa$ of the reference curve, $\Gamma_0$, being relieved by development of negative Gaussian curvature in the active mantle strip. The other interesting case is in Fig. \ref{fig:hierarchical}, where the positive Gaussian curvature after the first two growth increments, i.e., over $(t_0,t_1]$ and $(t_1,t_2]$ up to mantle configuration $\Omega^\text{m}_{t_{t_1}}$, changes into negative Gaussian curvature after the final volumetric growth increment in $\Omega^\text{m}_{t_{t_2}}$. Taking a cue from the temporal variation in volume growth over $(t_2,t_3]$, we recognize that it also imposes a spatial variation: the mantle strip of width $\delta s^\ast$ formed by surface growth over $(t_0,t_1]$ experiences volume growth $\delta g = \delta g^\ast$, but the strip of length $\delta s = \frac{1}{2}\delta s^\ast$ from surface growth over $(t_1,t_2]$ experiences a total volume growth of $\delta g = 3\delta g^\ast$. We are therefore prompted to consider that the rate of growth strain $\varepsilon_2$ is an increasing function along the $\bs_1$ direction, i.e., $\partial\varepsilon_2/\partial\xi_1 >0$, where $\xi_1$ is the curvilinear coordinate defining $\bs_1$. In this instance, within a single growth increment there is greater excess of length at the leading edge of the active mantle strip compared to the trailing edge. The elastic mantle attains a locally energy minimizing configuration by adopting negative Gaussian curvature of the deformed mantle.

\begin{figure}[!ht]
\begin{center} 
 \psfrag{u}{\small $\bs_2$}
 \psfrag{v}{\small $\bs_1$}
 \psfrag{w}{\small $\bs_3$}
 \psfrag{z}{\tiny $\bu=\bzero$}
\includegraphics[width=0.75\textwidth]{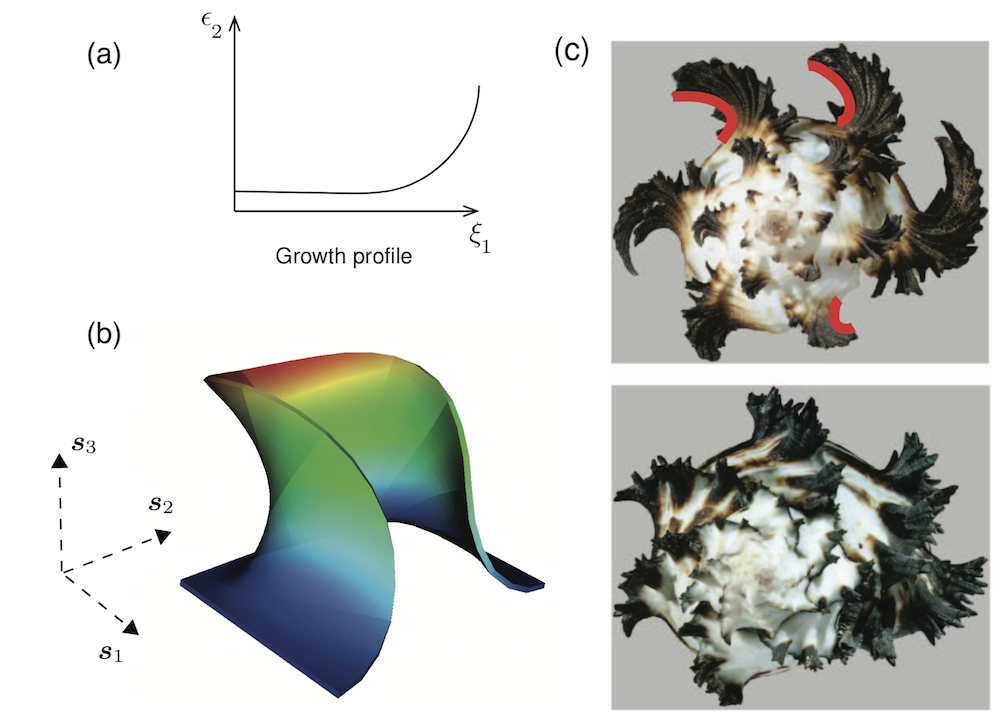}{}
\end{center}
\caption{{\bf Spatially varying volume growth} We impose volume growth strain increments that vary along the $\xi_1$ direction that is the curvilinear coordinate defining $\bs_1$, with an increasing gradient toward the leading edge, as shown in (a). The result appears in (b), displaying large, negative Gaussian curvature, mimicking the strongly backward arching morphology observed in a number of shell species, for example as seen also in (c) \emph{Hexaplex chicoreum}. Dirichlet boundary conditions, $\bu = \bzero$, are applied on the trailing surface (boundary) and the lateral surfaces (boundaries) of the mantle, which are perpendicular to $\bs_2$, and traction-free Neumann conditions, $\bP\bN = \bzero$, are applied on the front surface (boundary). Also see S6 Movie for the evolution of deformation leading to a morphology with strongly negative Gaussian curvature.}
\label{fig:foldback}
\end{figure}

An example of directly imposing such spatial variation is shown in Fig~\ref{fig:foldback}, where the profile of $\varepsilon_2(\xi_1)$ has low but positive curvature $\partial^2\varepsilon_2/\partial\xi_1^2 > 0$ for small $\xi_1$, changing smoothly to high curvature $\partial^2\varepsilon_2/\partial\xi_1^2 \gg 0$ for larger $\xi_1$. The variation in the rate of growth strain generates a deformation with a finite component in the negative $\bs_1$ direction, i.e. the mantle ``arches back'' to accommodate the excess length, creating ornamentation with negative Gaussian curvature.

While, as suggested by our computations and demonstrated in Fig. \ref{fig:arcMode} and Figs \ref{fig:growthRate}-\ref{fig:hierarchical},  antimarginal ornamentation in shells is often with Gaussian curvature that is positive or appears to vanish, there are a number of species of bivalves, cephalopods and gastropods that display such negative Gaussian curvature. In Fig~\ref{fig:foldback}(c) we show a top view of \emph{Hexaplex chicoreum}, which displays a strongly backward arching ornamentation, similar to the mantle deformation in Fig. \ref{fig:foldback}(b). Here we have another instance of a mechanical basis for a feature of ornamentation in mollusk shells for which no mechanistic explanation has previously been advanced, and that can be reproduced in our computational framework for coupled surface and volume growth. 

\section*{Discussion}
Mechanics has been recognized as a framework for explaining biological growth and form since at least the appearance of D'Arcy Thompson's work \cite{thompson1942growth}. However, a large part of the literature on morphological aspects of growth since the 1970s, such as that assembled by Meinhardt \cite{meinhardt1995algorithmic} and others, has focused on applying analytic or semi-analytic generating curves to the forms of shells, horns and antlers. The coupling of three-dimensional form to material forces and displacements, one aspect of which is \emph{morphoelasticity}, has remained a more difficult problem. The difficulty stems from the complexity attained by the coupled equations, especially where nonlinear elasticity appears, and has been very well laid out in \cite{amatar10} and \cite{goriely2017mathematics}. Consequently, it is only with the marriage of mathematics and numerical methods that general, three-dimensional, initial and boundary value problems have been solved \cite{gaargr04,gogeho15}. The literature on computational treatments of biological growth also has, in our eyes, suffered a limitation: problems addressable by the model of inhomogeneous, volume growth, i.e., \emph{morphoelasticity}, have formed the mainstay of this body of work. Effective as this treatment has been in explaining tumor growth \cite{narayanan2010silico, rudraraju2013multiphysics, mills2014}, aspects of cardiovascular systems, and the folding of soft, layered structures during morphogenesis \cite{bayly2014mechanical}, it cannot be elegantly extended to accretive, surface growth. For such problems, the morphoelastic treatment is restricted to representing advancing fronts by a thickening surface layer. Under its effect, the front's motion is an emergent phenomenon that is controlled by local, pointwise, volume growth. Neither the elaborate, generated surfaces, nor their accompanying elastic fields can be represented by such an application of volume growth with its foundations in local volume changes, rather than \emph{de novo} deposition of material.

Against this backdrop, we have presented, to the best of our knowledge, the first combined computational framework for accretive, surface growth and local, morphoelastic, volume growth. The mathematical basis for this framework in terms of generating surfaces, evolving reference configurations and moving fronts has been crucial because it has provided a rigorous foundation on which to elucidate the discretized, space-time formulation, as well as the finite element framework. The discretized space-time is a faithful reflection of the coupled processes of accretive surface growth and morphoelastic volume growth. There are alternatives to the finite element framework, however. Variants of level-set, phase field and immersed boundary methods would allow propagation of surface growth and the calcification front by fractions of an element width. Arbitrary changes in the propagation directions $\bs_1$ and $\bv^\text{c}$ could also be easily represented. We do not, however, see that this restriction to propagation by integral element widths presents a fundamental limitation in the shell morphologies and ornamentations that can be represented by the approach presented here. The advantages listed above for immersed boundary type methods could be approached by nonuniform element sizes in the advancing surface, and stepped fronts approximating changing directions on average.

\begin{figure}[!ht]
 \begin{center} 
 \psfrag{a}{$\delta g$}
 \psfrag{b}{$\delta \bs$}
 \psfrag{c}{Calcified shell}
 \psfrag{d}{$\Gamma_0$, curvature $\kappa$}
 \psfrag{e}{}
 \psfrag{u}{$\bs_2$}
 \psfrag{v}{$\bs_1$}
 \psfrag{w}{$\bs_3$}
 \psfrag{x}{$\delta s$}
 \psfrag{y}{$\delta g^{-1}$}
 \psfrag{z}{$\kappa$}
 \includegraphics[width=0.9\textwidth]{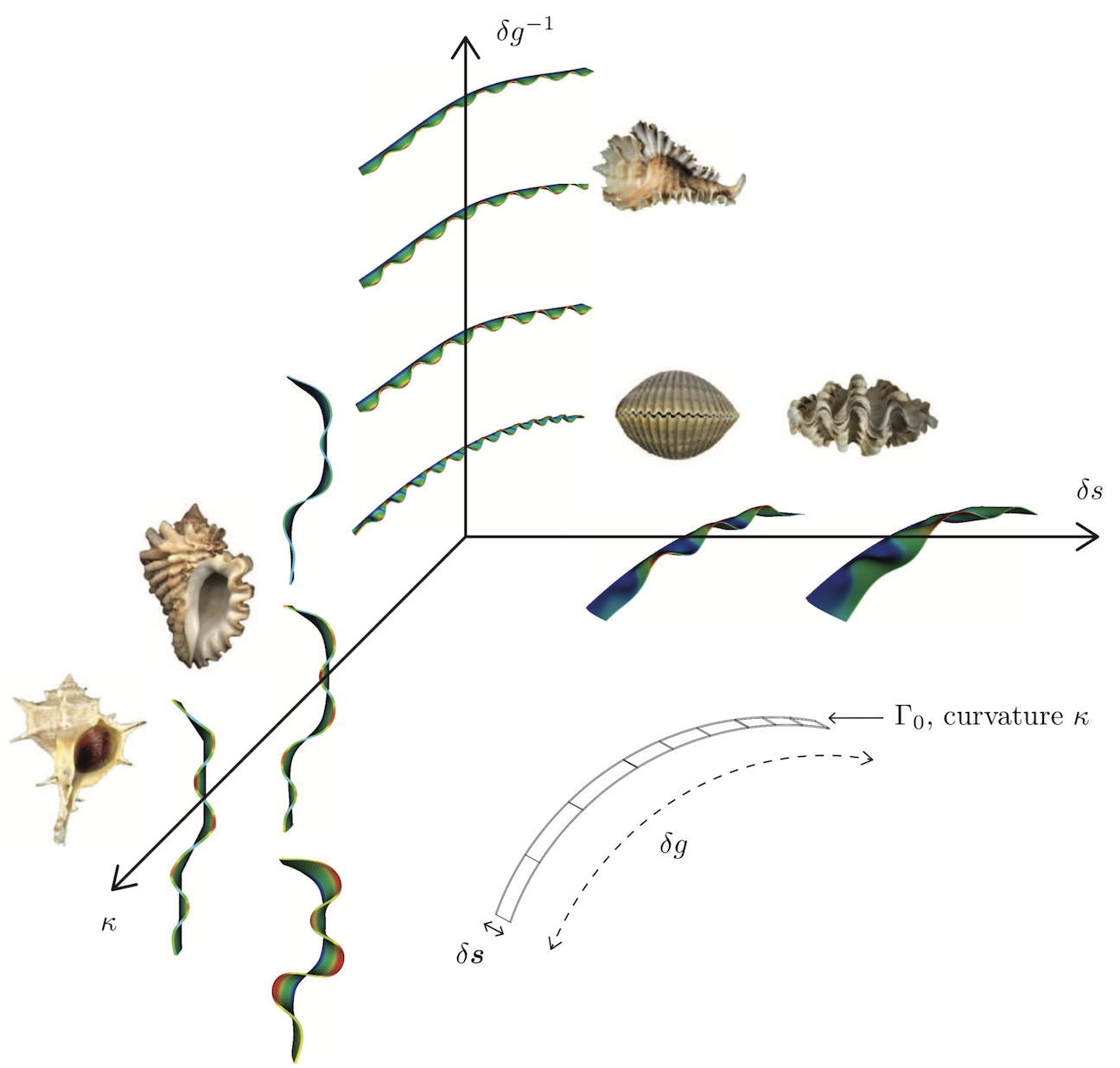}
 \caption{Phase diagram representing the effect of the growth and geometric parameters - growth strain increment ($\delta g$), active mantle width ($\delta s$), and curvature of the reference curve ($\kappa$), on the morphology of shell ornamentations.}
 \label{fig:phaseDiagram}
 \end{center}
 \end{figure}

Most crucially, our work has identified the prominent role played by geometry in controlling mantle deformation under the driver of morphoelastic growth. The active mantle width, which is a direct outcome of the surface growth rate, has a very visible influence on the mode number, mode shape and, as we have demonstrated, on the appearance of hierarchical ornamentation. The curvature of reference surfaces is the other parameter by which geometry acts directly to control the locations of crests and valleys. The evolving reference configurations, as each generation of active mantle is enslaved to the reference curvature of the previous generation, present a pathway for coupling of morphoelastic volume growth with surface growth and curvature. Figure \ref{fig:phaseDiagram} is a ``phase diagram'' illustrating salient aspects of our studies over this parameter space.

We have not attempted to compile computational demonstrations that match molluskan morphologies with high fidelity. While, in our opinion, most of our computations match well with features of actual molluskan morphologies, others such as in Figures \ref{fig:baseGeometry}c and \ref{fig:baseGeometry}d are less satisfying, especially in representation of spiny outgrowths (see the following paragraph in this regard). It is also true, however, that more complexity could be introduced to the model. Contact mechanics is one such addition, which is on our critical path, but must await a future communication.  Another is the spatial variation of material properties, which we have already addressed before (see following paragraph). While a detailed tuning over such effects may add insight to mechanisms, several trends are visible in what we have explored here. We have experimented with some variations on our basic themes; variations that typically are not describable with analytic forms, but are ubiquitous in nature, thus making them obvious candidates for computational exploration. This is the rational for investigations of temporal and spatial variations in growth.

We note that previous work by some of the authors \cite{moulton2012morphoelastic} already has highlighted the role of a spatial variation in material properties in shaping the sharp spines seen on the shells of bivalves, cephalopods and gastropods. The present communication adds to this emerging picture of the influence of physics, by shifting the focus to geometry, which, driven by morphoelastic growth, acts through the mechanism of surface growth and a parametric dependence on curvature.

Of particular interest is the further coupling of this framework with pattern formation by a range of reaction-transport equations. This would make accessible well-studied developmental milestones such as the patterning and morphogenesis of limbs and digits. Other problems in morphogenesis, such as the formation of skeletons are also within reach of our computational framework. Finally, we note that we have modelled antimarginal ornamentation events via bursts of growth. It is more likely that the growth rate does not change significantly at the location of ornamentations, but rather that the thickness of the mantle decreases, creating an increase in the length of the active mantle strip and a corresponding decrease in stiffness without requiring a strong increase in material. This effect would be interesting to incorporate in future studies, as the decrease in stiffness would further amplify the amplitudes of transverse deformation, and also may be non-uniform along the length of the mantle strip.

 \clearpage
 

\clearpage

\begin{flushleft}
{\Large
\textbf\newline{{\bf Supplementary Material} - A computational framework for the morpho-elastic development of molluskan shells by surface and volume growth} 
}
\newline
\\
Shiva Rudraraju \textsuperscript{1},
Derek E. Moulton \textsuperscript{2},
R\'egis Chirat \textsuperscript{3},
Alain Goriely\textsuperscript{2},
Krishna Garikipati \textsuperscript{4*},
\\
\bigskip
\textbf{1} Department of Mechanical Engineering, University of Wisconsin-Madison, Wisconsin, USA
\\
\textbf{2} Mathematical Institute, University of Oxford, Oxford, UK
\\
\textbf{3}  Université Lyon1, CNRS UMR 5276 LGL-TPE, France
\\
\textbf{4}  Departments of Mechanical Engineering and Mathematics, Michigan Institute for Computational Discovery \& Engineering, University of Michigan, Ann Arbor, Michigan, USA
\\
\end{flushleft}

\setcounter{equation}{0}
\setcounter{figure}{0}
\setcounter{table}{0}
\setcounter{page}{1}
\renewcommand{\theequation}{S\arabic{equation}}
\renewcommand{\thefigure}{S\arabic{figure}}

\section*{S1 Text. Buckling analysis of a plate}
We compute here the relationship between buckling mode and active mantle width via the buckling of a plate. We consider a plate with zero reference curvature of length $A$ in the $x$ direction and width $B$ in the $y$ direction (the $x$ and $y$ directions here correspond to the $s_2$ and $s_1$ directions in the main text, respectively, so we are primarily interested in the case $A\gg B$). The governing equation for the transverse deformation $w(x,y)$ of the plate is \cite{Timoshenko, Howell}
\begin{equation}\label{Eq:platePDE}
    D\nabla^4 w + N w_{xx}=0,
\end{equation}
where $D$ is the bending modulus and $N$ is a compressive force due to growth in the $x$ direction (defined as positive here). For boundary conditions, we take the plate to be clamped on one long edge, free on the other long edge, and simply supported on the two short edges. These conditions read
\begin{align}\label{Eq:plateBC}
    & w = 0, \;\; w_{xx}+\nu w_{yy} = 0 \;\; \text{on } x=0,\;A, \\
    & w = 0, \;\; w_y = 0 \;\; \text{on } y=0, \\
    & w_{yy}+\nu w_{xx} = 0, \;\;w_{yyy}+(2-\nu)w_{xxy} = 0 \;\; \text{on } y=B, \\
\end{align}
where $\nu$ is the Poisson ratio.
The system \eqref{Eq:platePDE}, \eqref{Eq:plateBC} has solution $$w(x,y)=\sin\left(\frac{m\pi x}{A}\right)f(y)$$ where $m$ is the buckling mode. Taking $f=\exp(i\lambda y)$ yields the characteristic equation
$$A^4 \lambda^4 - \frac{A^2 m^2}{\pi^2} (\tilde N - 2 \lambda^2) + m^4 \pi^4=0,$$
where $\tilde N=N/D$. This has roots
\begin{align}
& \lambda_1^{\pm}=\pm i\frac{\sqrt{A m \sqrt{\tilde N}\pi + m^2\pi^2}}{A},\\
& \lambda_2^{\pm}=\pm \frac{\sqrt{A m \sqrt{\tilde N}\pi - m^2\pi^2}}{A},
\end{align}
a nontrivial solution only existing if $\tilde N>m^2\pi^2/A^2$. The function $f(y)$ thus takes the form
$$f(y)=c_1\exp(i\lambda_1^+ y)+c_2\exp(i\lambda_1^- y)+c_3\cos(\lambda_2^+ y)+c_4\sin(\lambda_2^+ y).$$
The boundary conditions in the $y$-direction translate to 
$$f(0)=f'(0)=0,\;\;f''(B)-\nu\frac{m^2\pi^2}{A^2}f(B)=f'''(B)-(2+\nu)\frac{m^2\pi^2}{A^2}f'(B)=0.$$ Imposing these conditions yields an eigenvalue problem for the critical compression $\tilde N=\tilde N^*$, and the critical buckling mode is determined by finding the integer value of $m$ at which $\tilde N^*$ is minimized. The points in Fig 8 of the main text were produced by fixing $A=20$, $\nu=0.3$, varying $B$ between 0.5 and 2.5, and computing the critical mode at each width.

\begin{figure}[h]
\begin{center}
\includegraphics[width=0.75\textwidth]{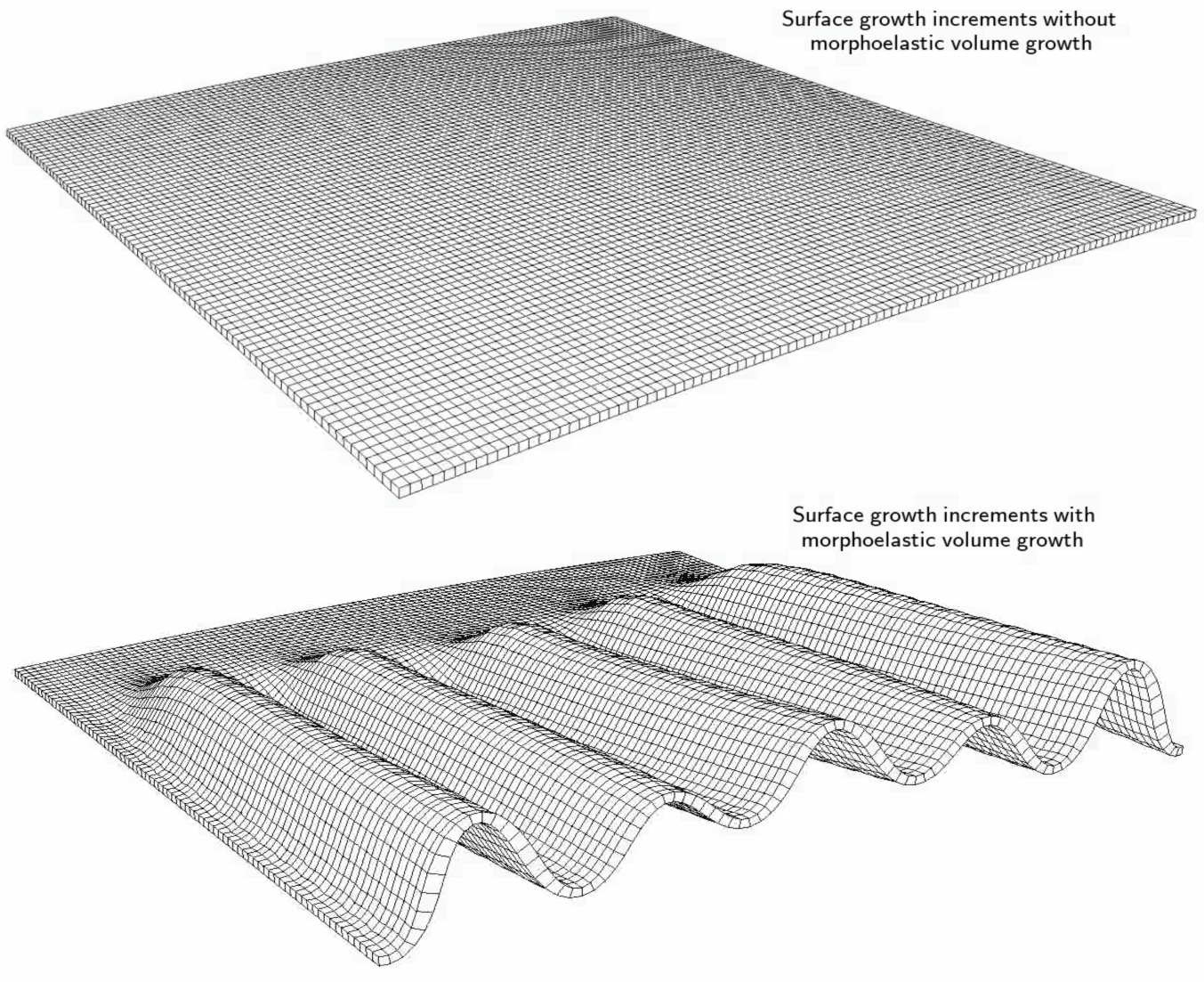}
\caption*{{\bf S1 Movie. Space-time discretization of the molluskan shell in the finite element framework}. Shown at the top is the evolution of the geometry (finite element mesh) for 20 surface growth increments (addition of the mantle in 20 incremental strips, each being four elements wide) of a representative molluskan shell in its reference configuration, without the morphoelastic volume growth increments. As a result the reference configuration remains a flat plate. The accompanying computation at the bottom shows the 20 growth increments with the complete model (surface growth, volume growth and calcification of 20 mantle strips in sequence). Beginning with a flat plate geometry, each surface growth increment is followed by its morphoelastic volume growth increment and calcification.Calcification is the final stage of the sequence for each such mantle strip of four elements. Therefore, at any instant, it is only the mantle strip at the leading edge that undergoes morphoelastic volume growth. The mantle strips that grew before it have already undergone calcification. This is the case for S1 Movie-S3 Movie, and S5 Movie.}
\label{fig:suppFig1} 
\end{center}
\end{figure}
 
\begin{figure}[h]
\begin{center}
\includegraphics[width=0.3\textwidth]{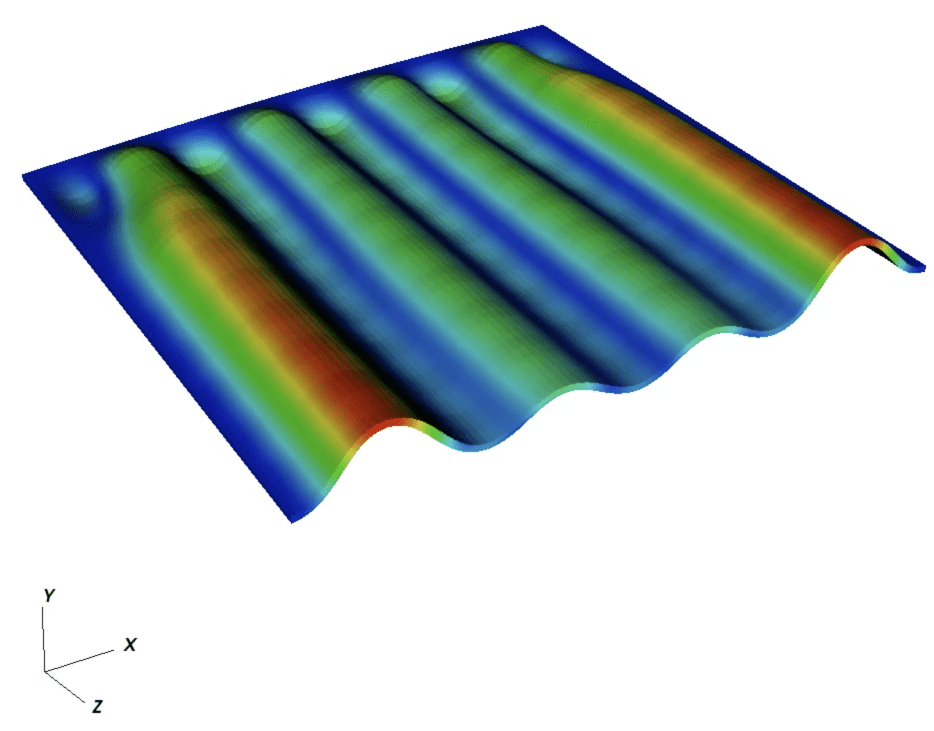}
\caption*{{\bf S2 Movie. Influence of the geometry of reference curves on antimarginal ornamentation: Planar geometry.} Shown is the evolution of 10 mantle strips starting from a flat plate geometry of the reference curve. The contour colors indicate the normalized displacement magnitude. As in S1 Movie, at any instant, it is only the mantle strip at the leading edge that undergoes morphoelastic volume growth. The mantle strips that grew before it have already undergone calcification.}
\label{fig:suppFig2} 
\end{center}
\end{figure}

\begin{figure}[h]
\begin{center}
\includegraphics[width=0.4\textwidth]{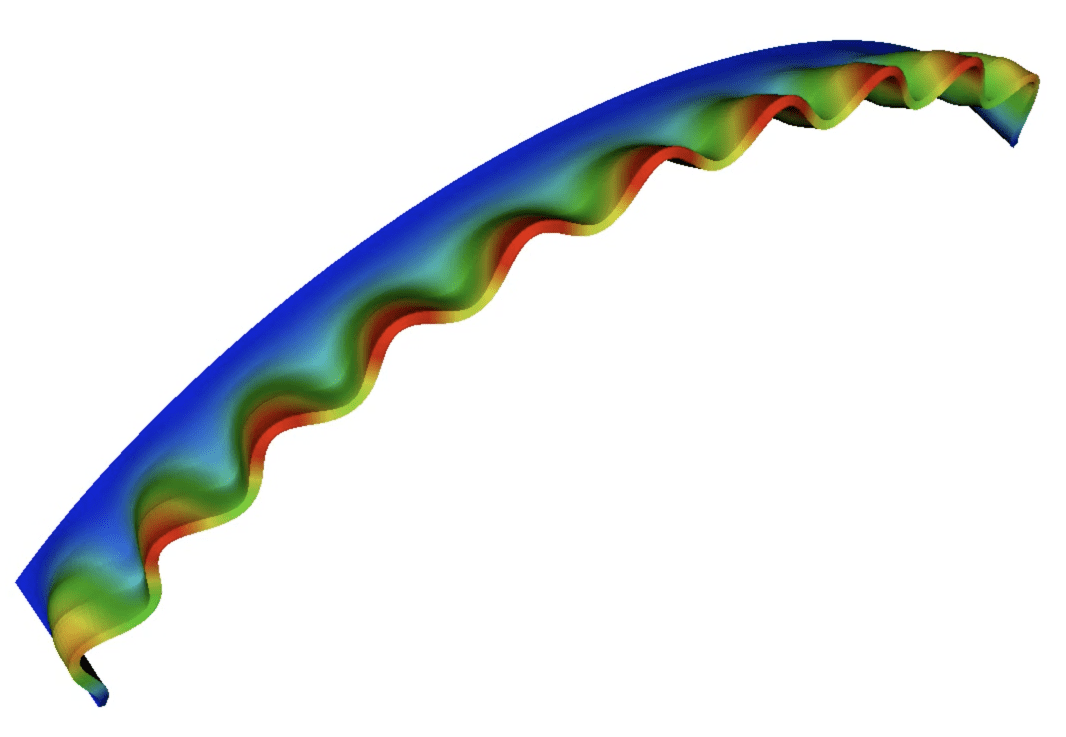}
\caption*{{\bf S3 Movie. Influence of the geometry of reference curves on antimarginal ornamentation: Arc geometry.} Shown is the evolution of 4 mantle strips starting from an arc geometry of the reference curve. The contour colors indicate the normalized displacement magnitude. As in S1 Movie and S2 Movie, at any instant, it is only the mantle strip at the leading edge that undergoes morphoelastic volume growth. The mantle strips that grew before it have already undergone calcification.}
\label{fig:suppFig3} 
\end{center}
\end{figure}

\begin{figure}[h]
\begin{center}
\includegraphics[width=0.15\textwidth]{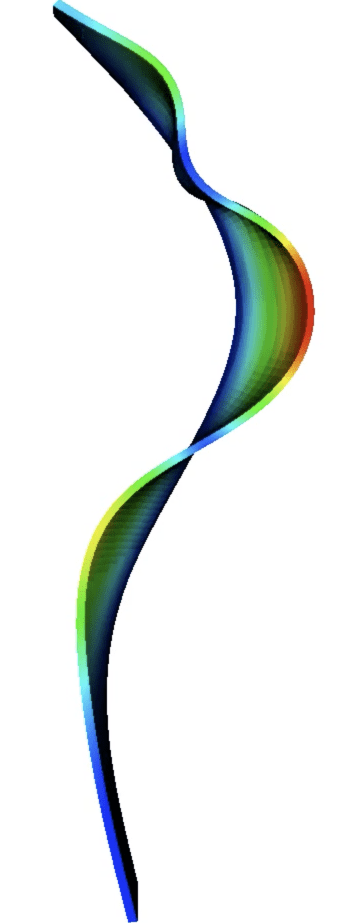}
\caption*{{\bf S4 Movie. Influence of the geometry of reference curves on antimarginal ornamentation: An arc with positive and negative curvature.} Shown is the evolution of a single mantle strip starting from a reference curve that is an arc with curvature of changing signs. Note that there is no surface growth in this movie. The trailing edge is the the only calcified part of the shell. The snap-through events seen during the deformation of the mantle strip are the elastic bifurcations (buckling modes) triggered by growth over this geometry. There are several bifurcations occurring in rapid succession due to volume growth within a single growth increment, in this simulation. Because of the stiff, nonlinear response of the shell undergoing elastic bifurcations, the single increment of volume growth is numerically applied as 20 load steps, and the evolution of the deformed geometry after each load step is shown in the corresponding evolution of the 20 frames shown in this movie. The contour colors indicate the normalized displacement magnitude.}
\label{fig:suppFig4} 
\end{center}
\end{figure}

\begin{figure}[h]
\begin{center}
\includegraphics[width=0.45\textwidth]{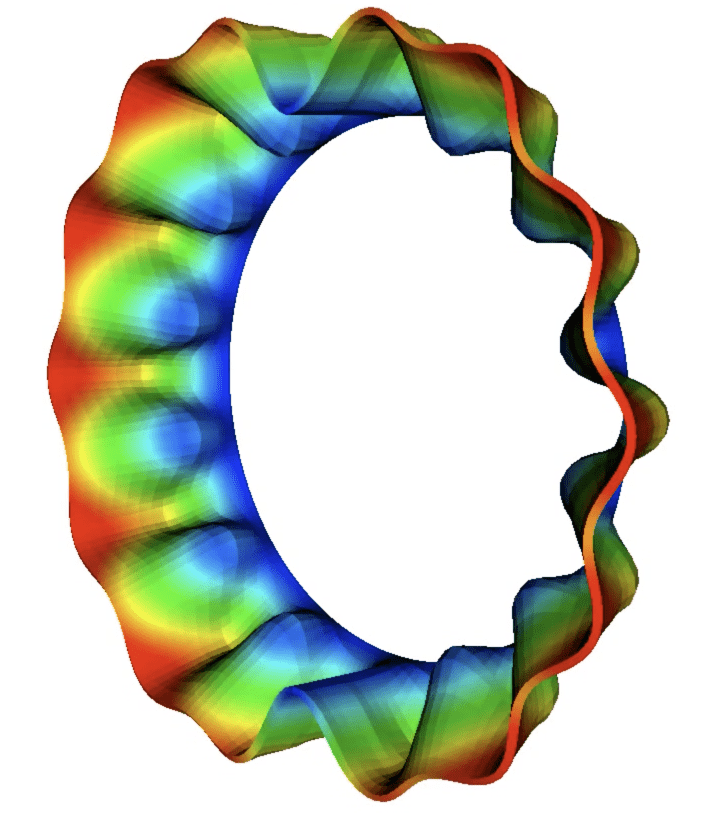}
\caption*{{\bf S5 Movie. Influence of the geometry of reference curves on antimarginal ornamentation: Closed circular geometry.} Shown is the evolution of 3 mantle strips starting from a circular geometry of the reference curve, and is representative of growth over a closed shell geometry. The contour colors indicate the normalized displacement magnitude. As in S1 Movie-S3 Movie, at any instant, it is only the mantle strip at the leading edge that undergoes morphoelastic volume growth. The mantle strips that grew before it have already undergone calcification.}
\label{fig:suppFig5} 
\end{center}
\end{figure}

\begin{figure}[h]
\begin{center}
\includegraphics[width=0.45\textwidth]{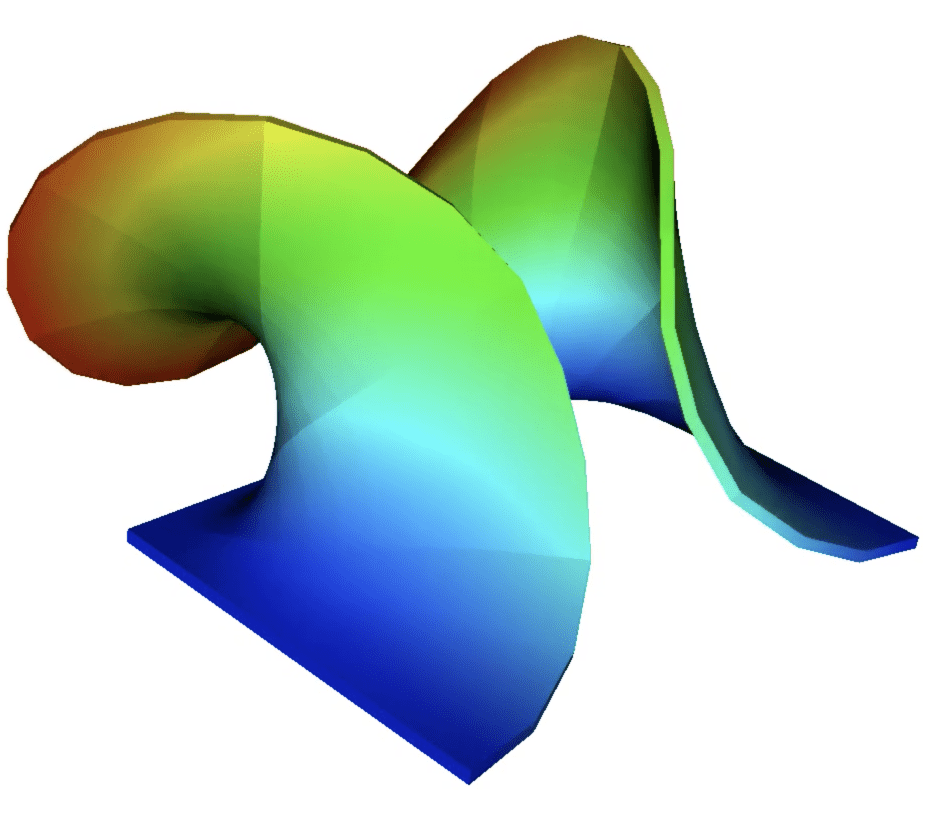}
\caption*{{\bf S6 Movie. Evolution of deformation leading to a backward arching morphology.} The contour colors indicate the normalized displacement magnitude. Note that there is no surface growth in this movie. The trailing edge is the only calcified part of the shell.}
\label{fig:suppFig6} 
\end{center}
\end{figure}
 
\end{document}